\documentclass[reprint,superscriptaddress,amsmath,twocolumn]{revtex4-2}

\usepackage{dcolumn}
\usepackage{bm}
\usepackage[charter,greekuppercase=italicized]{mathdesign}
\usepackage{microtype}
\usepackage{array}
\usepackage[usenames,dvipsnames]{xcolor}
\definecolor{nblue}{rgb}{0.0, 0.0, 1.0}
\definecolor{blue}{rgb}{0.0, 0.0, 1.0}
\definecolor{magenta}{rgb}{0.79, 0.08, 0.48}
\usepackage[colorlinks,linkcolor=nblue,urlcolor=magenta,citecolor=magenta,plainpages=false,pdfpagelabels,breaklinks]{hyperref}

\newcommand{\beq}{\begin{equation}}
\newcommand{\eeq}{\end{equation}}
\newcommand{\bea}{\begin{eqnarray}}
\newcommand{\eea}{\end{eqnarray}}

\newcommand{\tcb}[1]{{\color[rgb]{0,0,0}{#1}}}

\definecolor{colorGian}{RGB}{0,0,0} 
\newcommand{\comm}[1]{\textcolor{colorGian}{#1}}

\usepackage[charter,greekuppercase=italicized]{mathdesign}
\usepackage{graphicx,color,booktabs,microtype,afterpage} 
\graphicspath{{./}{figures/}}

\newcommand{\zfmu}{ZF-{\textmu}SR}
\newcommand{\tfmu}{TF-{\textmu}SR}

\newcommand{\musr}{{\textmu}SR}
\newcommand{\vga}{V$_2$Ga$_5$}

\renewcommand{\figurename}{Fig.}
\renewcommand{\tablename}{Table}
\makeatletter\renewcommand{\fnum@figure}[1]{\textbf{\figurename~\thefigure\,\textbar\,}}\makeatother
\makeatletter\renewcommand{\fnum@table}[1]{\tablename~\thetable\,\textbar\,}\makeatother

\begin{document}



\title{Anisotropic superconductivity in the quasi-one-dimensional superconductor V$_2$Ga$_5$}
\author{G.\ Lamura}
\affiliation{CNR-SPIN, I-16152 Genova, Italy} 
\author{D.\ Tay}
\affiliation{Laboratorium f\"ur Festk\"orperphysik, ETH Z\"urich, CH-8093 Z\"urich, Switzerland}
\author{R.\ Khasanov}
\affiliation{PSI Center for Neutron and Muon Sciences CNM, CH-5232 Villigen PSI,  Switzerland}
\author{P.~Gentile}
\affiliation{SPIN-CNR, IT-84084 Fisciano (SA), Italy, c/o Università di Salerno, IT-84084 Fisciano (SA), Italy}
\author{C.\ Q.\ Xu} 
\affiliation{School of Physical Science and Technology, Ningbo University, Ningbo 315211, China}
\author{X.\ Ke} 
\affiliation{Department of Physics and Astronomy, Michigan State University, East Lansing, Michigan 48824-2320, USA}
\author{I. J. Onuorah} 
\affiliation{Dipartimento di Scienze Matematiche, Fisiche e Informatiche, Università di Parma, Parco Area delle Scienze 7/A, I-43124 Parma, Italy}
\author{P. Bonfà} 
\affiliation{Department of Physics, Informatics and Mathematics, University of Modena and Reggio Emilia, via Campi 213/a, 41125 Modena, Italy}
\affiliation{CNR-NANO S3, Istituto Nanoscienze, Via Campi 213/a, 41125 Modena, Italy}
\author{Xiaofeng Xu} \email[Corresponding author: ]{xuxiaofeng@zjut.edu.cn}
\affiliation{
School of Physics, Zhejiang University of Technology, Hangzhou 310023, China}
\author{T.\ Shiroka} \email[Corresponding author: ]{toni.shiroka@psi.ch}
\affiliation{Laboratorium f\"ur Festk\"orperphysik, ETH Z\"urich, CH-8093 Z\"urich, Switzerland}
\affiliation{PSI Center for Neutron and Muon Sciences CNM, CH-5232 Villigen PSI,  Switzerland}

\date{\today}

\begin{abstract}
The intermetallic quasi-one-dimensional binary superconductor \vga\ was
recently found to exhibit a topologically nontrivial normal state, making
it a natural candidate for a topological superconductor (TSC).
By combining dc-magnetization, nuclear magnetic resonance (NMR), and
muon-spin rotation (\musr) measurements on high-quality \vga\ single
crystals, we investigate the electronic properties of its normal- and
superconducting (SC) ground states. NMR measurements in the normal state
indicate a strong anisotropy in both the line shifts and the relaxation
rates. Such anisotropy persists also in the superconducting state, as
shown by the magnetization- and \musr-spectroscopy results. In the latter
case, data collected at different temperatures, pressures, and
directions of the magnetic field evidence a fully-gapped, strongly anisotropic superconductivity.
At the same time, hydrostatic pressure is shown to only lower the $T_c$
value, but not to change the superfluid density nor its temperature
dependence. Lastly, we discuss the search for topological signatures in
the normal state of \vga, as well as a peak splitting in the FFT of the
{\textmu}SR spectrum, possibly related to an unconventional vortex lattice.
Our results suggest that \vga\ is a novel system, whose anisotropy plays
a key role in determining its unusual electronic properties.
\end{abstract}

\maketitle

Following the advent of topological insulators, the search for quantum
materials with symmetry-enforced topological states has attracted
widespread research interest in both the condensed-matter- and the
material-science communities~\cite{Hasan2010,Qi2011,Li2022}.
In this new class of materials, the systems that simultaneously exhibit
superconductivity and topologically nontrivial electronic bands, with
Weyl- and/or Dirac nodes, are the most promising candidates for the
realization of topological and/or nematic superconductivity.
The potential to host Majorana fermions in their vortex cores represents
a significant advantage in the pursuit of quantum computing.

\vga, currently known as a candidate TSC material, was discovered already
in the 1960s. Back then, in the search for new superconducting alloys,
attempts to grow V$_3$Ga by cooling a gallium-rich vanadium solution
resulted in a new compound, \vga\ ~\cite{VanVucht1963,Cruce1974,Lobring2002}.
\vga\ crystallizes in the form of long needles, which exhibit a unique
quasi-1D structure, as confirmed by x-ray diffraction measurements
and band-structure calculations on single crystals~\cite{Lobring2002}.
Here, one-dimensional vanadium chains along the $c$ axis are coupled
to each other by neighboring Ga ions. This pronounced 1D character
makes \vga\ a natural platform to study the effects of strong crystalline
anisotropy on the electronic properties and the possible occurrence of
topological features. However, as its superconducting
temperature is only 3.6\,K, \vga\ was forgotten for a long time.

The increased interest in topological quantum computing has re-ignited
interest in the TSCs candidate materials, including \vga. Recent studies,
comprising mostly bulk techniques, such as heat-capacity and transport
measurements, as well as some preliminary \musr\ results~\cite{Xiaofeng2024,Cheng2024},
generally agree that \vga\ possesses multiple nodeless superconducting gaps.
However, \tfmu\ measurements in a magnetic field orthogonal to
the $c$-axis~\cite{Cheng2024} show a superconducting shielding fraction
of only 60\% and a rather weak depolarization rate (of $\sim 1$\,{\textmu}s$^{-1}$)~\cite{Cheng2024}, which casts serious doubts on the intrinsic nature of the
extracted electronic properties.

In this work, we present an extensive, more detailed investigation,
including bulk dc-magnetization, NMR, and \musr, all performed with the
applied magnetic field parallel ($H^{\parallel}$) and perpendicular
($H^{\perp}$) to the \textit{c}-axis, as well as \musr\ measurements
under hydrostatic pressure on aligned single crystals. 
All the crystals used in our measurements possess an \textit{excellent
crystalline quality}, as confirmed by a shielding volume fraction of
about 100\% (measured via dc-susceptibility) and a superconducting volume
fraction of $\sim 95$\% (determined by \musr\ spectroscopy). In particular,
in the latter case, the low-$T$ muon depolarization rate
($\sim 2.7$\,{\textmu}s$^{-1}$ at 0.3\,K) is almost three times larger than that measured on \vga\ single crystals under the same conditions~\cite{Cheng2024}.

The improved quality of samples is critical for investigating
their intrinsic superconducting properties, as well as for obtaining
new insight on this unique compound. Our NMR and {\textmu}SR in the
normal- and the superconducting state reveal a strong anisotropy of
in-plane vs.\ out-of-plane electronic properties. At the same time, 
transverse-field {\textmu}SR in the superconducting state shows an
unexpected peak splitting in the Fourier transform of the time-dependent
asymmetry, hinting at the emergence of an unconventional
vortex lattice that has never been observed in any other topological
material. 

\begin{figure*}[!thp]
	\centering
	\vspace{-1ex}%
	\centering
\includegraphics[width=.95\linewidth]{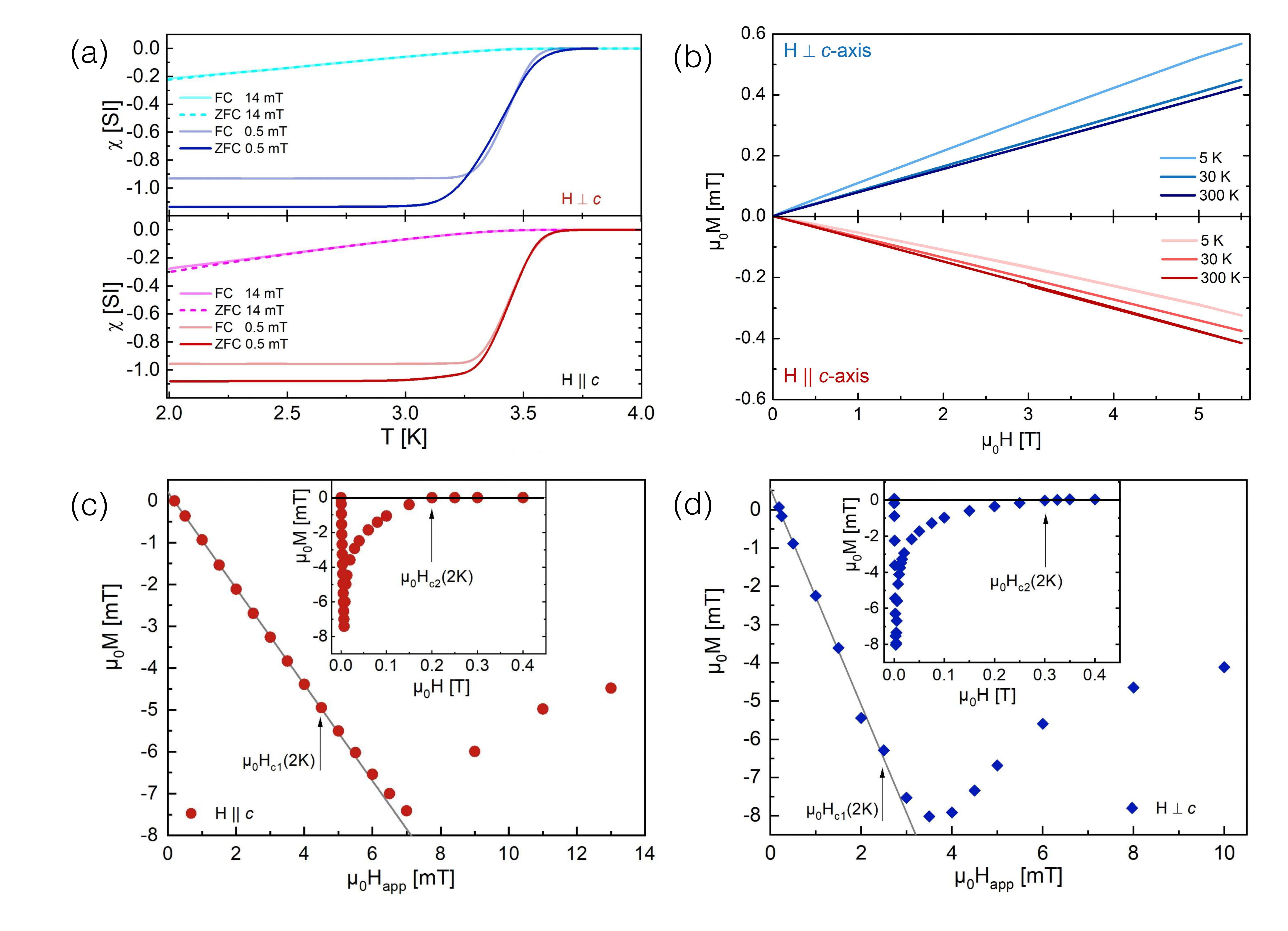}
	\caption{\label{fig:Hc1}(a) Temperature-dependent magnetic susceptibility
	of \vga, measured in an applied field of 0.5 and 14\,mT, using the ZFC-
	and FC protocols with the magnetic field applied parallel and perpendicular
	to the $c$-axis. (b) Normal-state magnetization vs magnetic field.
	Note the opposite sign of the magnetic response in the two cases.
	Field-dependent magnetization $M(H)$ recorded below $T_c$ (here, at 2\,K)
	for the field applied parallel- (c) and perpendicular (d) to the $c$-axis. The vertical arrows indicate the estimated lower and the upper critical fields at 2 K (see text for details).}
\end{figure*}
%

\vspace{15pt}
\noindent{\large\textbf{Results}}\\[0.3ex]
Since \vga\ crystals are needle-shaped, all the measurements had
to be carried out on groups of aligned single crystals (bundles). It is
important to note that, while the whole bundle behaves as a single crystal when the applied field is parallel to the \textit{c}-axis ($H^{\parallel}$ configuration), it behaves as a polycrystal when the field is perpendicular to it ($H^{\perp}$ configuration).
This is due to the random orientation of the \textit{a}-axis of each
crystal within the bundle. Further details about the geometry of
the bundles are provided in the Supplementary Information.\\[0.8ex]
\noindent\textbf{DC magnetization measurements}\\[0.3ex]
Figure~\ref{fig:Hc1}a shows the low-field intrinsic susceptibility
$\chi$ as a function of temperature, measured at 0.5 and 14\,mT.
Due to sample-shape demagnetization effects, the measured susceptibility
$\chi_\mathrm{meas}$ generally differs from the intrinsic one, the
relation between the two being $\chi_\mathrm{meas} = \chi/(1 + N \chi)$,
where $N$ is the demagnetization factor. Our needle-like \vga\ samples
can be considered as long cylinders, with a length-to-diameter ratio
$l/d > 10$. In this case, the demagnetization factor is
$N^{-1} = 2 + (1/\sqrt{2})\,d/l$~\cite{Prozorov2018}, and it can be taken
equal to 0 and 0.5 for the parallel and perpendicular configuration, respectively (see also Table VI in Ref.~\cite{Chen1991}). 
%
%
DC sus\-cep\-ti\-bil\-i\-ty results indicate that:
(i) the full shielding at the lowest field is suggestive of bulk superconductivity
and (ii) the small difference between the zero-field cooling (ZFC) and
field cooling (FC) datasets is reminiscent of a negligible vortex pinning,
which is even smaller in the $H^{\parallel}$ configuration (as confirmed
by our \musr\ measurements --- see below). 

In Fig.~\ref{fig:Hc1}b we show a representative set of isothermal
magnetization curves recorded in the normal state. Interestingly, the
field response is paramagnetic (diamagnetic ~\footnote{\tcb{\tfmu\ results indicate the
presence of an impure phase of about 5\%, most likely paramagnetic.
As such, it could be responsible for the slight increase in dc magnetization
at low temperature, instead of the expected $T$-independent diamagnetic contribution.
The same effect could also arise because of a slight sample misalignment
during the dc susceptibility measurements.}})
for magnetic fields applied perpendicular (parallel) to
the $c$-axis. The presence of diamagnetism in the normal state of
\vga\ is consistent with the diamagnetism found in the normal state of
elemental Ga~\cite{Pankey1960}, as well as in the binary Ga-based intermetallic superconductors, such as Mo$_8$Ga$_{41}$~\cite{Verchenko2016} and Mo$_4$Ga$_{21}$~\cite{Verchenko2020}. On the other hand, to the best of our knowledge, \vga\ is one of the few compounds to exhibit a significant magnetic anisotropy in the normal state: paramagnetic vs.\ diamagnetic for applied fields perpendicular- or parallel to the $c$-axis, respectively. A similar behavior has been observed in single crystals of antimony-tin alloys~\cite{Hart1937,Lonsdale1937}. Such \comm{unusual magnetic anisotropy} was attributed to the Stark effect, because the crystal electric field is likely to quench the spin susceptibility~\cite{Lonsdale1937} in favor of the orbital diamagnetic contribution. This may also be the case for \vga, where both Ga and V possess sizable quadrupolar moments. This picture is supported by the NMR and {\textmu}SR results presented in the following sections. 

In Fig.~\ref{fig:Hc1}c and \ref{fig:Hc1}d we plot the ZFC magnetization $M$ vs.\ the applied
magnetic field $H_\mathrm{app}$ at 2\,K. 
At this tem\-per\-a\-ture, we note that $\mu_{0}H_{c2}^{\parallel} \sim 0.2$\,T and
$\mu_{0}H_{c2}^{\perp} \sim 0.3$\,T. As for the lower critical field
$H_{c1}$, this is defined as the threshold value beyond which $M$
starts to deviate from the linearity (Meissner effect).
The resulting critical fields (at 2\,K) are
$\mu_{0}H_{c1}^{\parallel} \sim 4.5$\,mT and
$\mu_{0}H_{c1}^{\perp}     \sim 2.5$\,mT, respectively.
We remark that both the lower- and the upper critical field
(here measured at 2\,K) are in good agreement with
the values reported in literature at the same temperature~\cite{Xiaofeng2024,Cruce1974,Cheng2024}. 
Finally, by taking into account that the internal magnetic field
is $H_\mathrm{int} = H_\mathrm{app} - NM$ and the above mentioned values
for $N$, the slopes of the linear low-field regions allow us to
estimate the shielding volume fractions at 2\,K. These are
$\chi^{\parallel} \sim -1.150(4)$ and $\chi^{\perp} \sim -1.17(2)$,
respectively, in good agreement with the low-$T$ values resulting
from the dc magnetic susceptibility taken at 0.5 mT (Fig.~\ref{fig:Hc1}a).\\

\begin{figure*}[t]
	\centering
\includegraphics[width=0.95\linewidth]{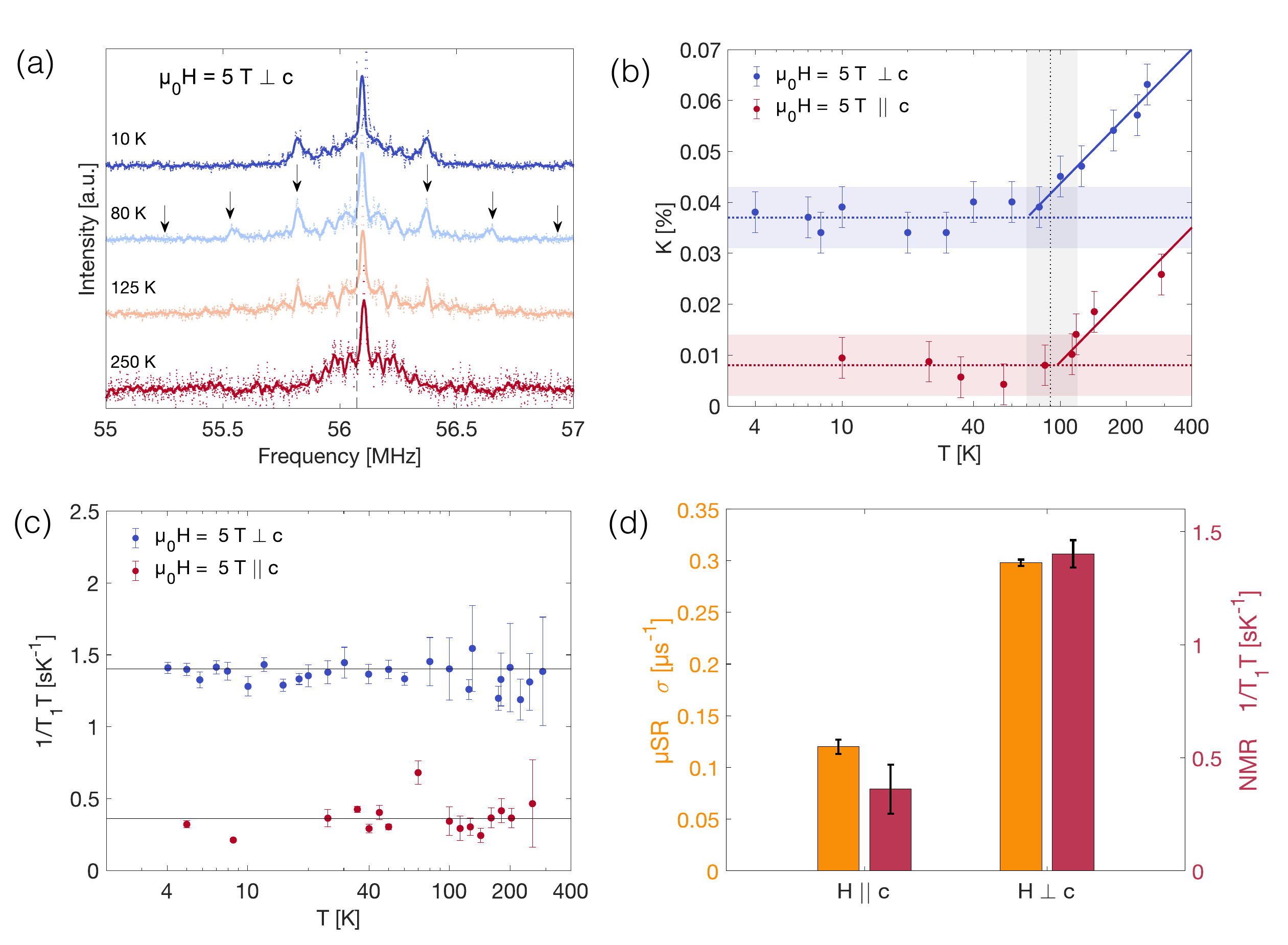}
	\caption{\label{fig:NMR_results}(a) $^{51}$V NMR lineshapes at 5\,T.
	The presence of satellites (shown by arrows for the 80\,K case)
	indicates a strong quadrupole interaction, 
	in turn arising from the strong anisotropy of the \vga\ unit cell. 
	(b) Temperature dependence of the total NMR shift in \vga, which includes contributions from both the magnetic- and quadrupole interactions. While generally temperature-independent, above 90\,K, it shows a slight upward shift.
	The solid line indicates a fit to $K = a + b\ln(T)$. (c) Temperature
	dependence of the Korringa product $1/T_1T$. Its practically constant
	value in the normal state indicates an ideal-metal behavior.
	d) Anisotropy of relaxation measured via NMR (magenta) and {\textmu}SR
	(orange) (see Fig.~\ref{fig:TF_muSR}). Both techniques clearly
	indicate that \vga\ relaxes much faster in the $H^{\perp}$ configuration.}
\end{figure*}

\noindent\textbf{\textcolor{black}{\label{ssec:NMR}NMR in the normal state}}

We performed NMR measurements in a magnetic field of 5\,T, oriented
parallel and perpendicular to the \vga\ $c$-axis. To achieve the
H$^{\parallel}$ configuration, a special sample configuration had
to be realized (see Supplementary Information). 
The main results, including the line shapes at selected temperatures
and the evolution of the NMR shift and of $1/(T_1T)$ with
temperature are shown in Fig.~\ref{fig:NMR_results}a-c, respectively.
We evidence here three notable aspects.

Firstly, the $^{51}$V NMR line shapes in the $H^{\perp}$ configuration
indicate that, aside from the main NMR resonance peak at 56.1\,MHz,
there is also clear evidence of additional peaks. At 80\,K, where the
signal-to-noise ratio is the highest, five equally-spaced peaks can be
clearly distinguished. Since \vga\ has only one crystallographically
unique $^{51}$V site~\cite{Lobring2002}, the side peaks cannot be due
to multiple $^{51}$V sites. Instead, most likely, these additional peaks
arise from the quadrupolar satellite transitions, with $2\nu_Q = 274\pm 2$\,kHz. 
Since $^{51}$V is a spin 7/2 nucleus, one would expect to observe three
quadrupole satellites on each side of the main peak. However, the outermost
peaks are most likely hidden in the noise floor. In any case, the presence of
quadrupolar satellites in $^{51}$V NMR indicate the
presence of a strong local electric field gradient (EFG), a key signature of the anisotropy
of the electronic properties of \vga. Furthermore, the non-negligible
quadrupole coupling observed via NMR in the normal state is consistent
with our key result of anisotropic superconductivity in the
superconducting state.

Secondly, the line shifts reported in Fig.~\ref{fig:NMR_results}b and
the Korringa product ($1/T_{1}T$) in Fig.~\ref{fig:NMR_results}c are
both largely temperature independent, except for a small increase in NMR shift above 100\,K, almost negligible compared to the relevant error bars. 
Despite a lack of changes with temperature, the relaxation
rates and line shifts are still clearly different in the $H^{\parallel}$
and $H^{\perp}$ configurations. This confirms the normal-state
anisotropic behavior already revealed by the dc susceptibility results. 
A temperature-independent Korringa product and NMR shift indicate a
temperature-independent density of states (DOS)~\cite{Korringa1950}.
This is expected, considering that the NMR measurements were conducted
in the normal phase of a metal. We also note that previous
calculations~\cite{Lobring2002} show that the total DOS of \vga\ is 
8 (V) + 3 (Ga1) + 0.7 (Ga2) = 11.7 states/(eV f.u.). Considering that
pure vanadium has a DOS of approximately 2 states/(eV f.u.), as calculated
via different methods~\cite{Sihi2020}, one would expect the relaxation
rate (and, hence, the Korringa product) to be much faster in \vga.
Instead, we observe the opposite result, i.e., the Korringa product of
\vga\ in the parallel field configuration, 1.40(6)\,sK$^{-1}$, is almost
half of that measured in pure vanadium (3\,sK$^{-1}$)~\cite{Noer1964}.
The Korringa product measured in the perpendicular configuration is even
lower, at 0.4(1)\,sK$^{-1}$. Further investigation is needed to
understand the surprisingly short relaxation rate observed in \vga.

Thirdly, we note that the NMR shifts reported in 
Fig.~\ref{fig:NMR_results}b and the relaxation rates reported in
Fig.~\ref{fig:NMR_results}c are both highly anisotropic. We first
consider the anisotropy in the NMR line shift. Since, in the
$H^{\perp}$ configuration, the dc-magnetization is positive
(paramagnetic), while, in the $H^{\parallel}$ configuration, it is negative
(diamagnetic) (see Fig.~\ref{fig:Hc1}b), we expect the NMR shift,
which is a measure of spin susceptibility, to be more negative in the
$H^{\parallel}$ configuration. Furthermore, it is known that the NMR
frequency shift of the central transition in highly anisotropic
systems includes a large second-order quadrupolar shift term.  For
example, in the quasi-1D compound $\beta$-Sr$_{0:33}$V$_2$O$_5$
\cite{Waki2007}, the strong quadrupole interaction term results in an
additional negative $^{51}$V NMR shift of $\approx$ 0.5$\%$.  Hence, the combination of magnetic and quadrupolar effects results in a strong anisotropy in the NMR shifts measured in the $H^{\parallel}$ and H$^{\perp}$ configurations. In the case of the NMR relaxation rates, the temperature-independent NMR
Korringa product ($1/T_{1}T$) is 1.40(6)\,sK$^{-1}$ in the $H^{\perp}$
configuration, but it drops by a factor of 3.5, i.e., to 0.4(1)$^{-1}$\,sK$^{-1}$
in the $H^{\parallel}$  configuration. Such drop is consistent
with strong decrease in the {\textmu}SR relaxation rate (measured at 5\,K)
from 0.298(3)\,{\textmu}s$^{-1}$ in the $H^{\perp}$ configuration to
0.120(7)\,{\textmu}s$^{-1}$ in the $H^{\parallel}$ configuration.
This is a strong indication that the anisotropy in the NMR- and
{\textmu}SR measurements has the same structural origin, here arising
from the 1D vanadium chains.
\newpage

\begin{figure*}[t]
	\centering
	\vspace{-1ex}%
	\centering
\includegraphics[width=0.9\linewidth]{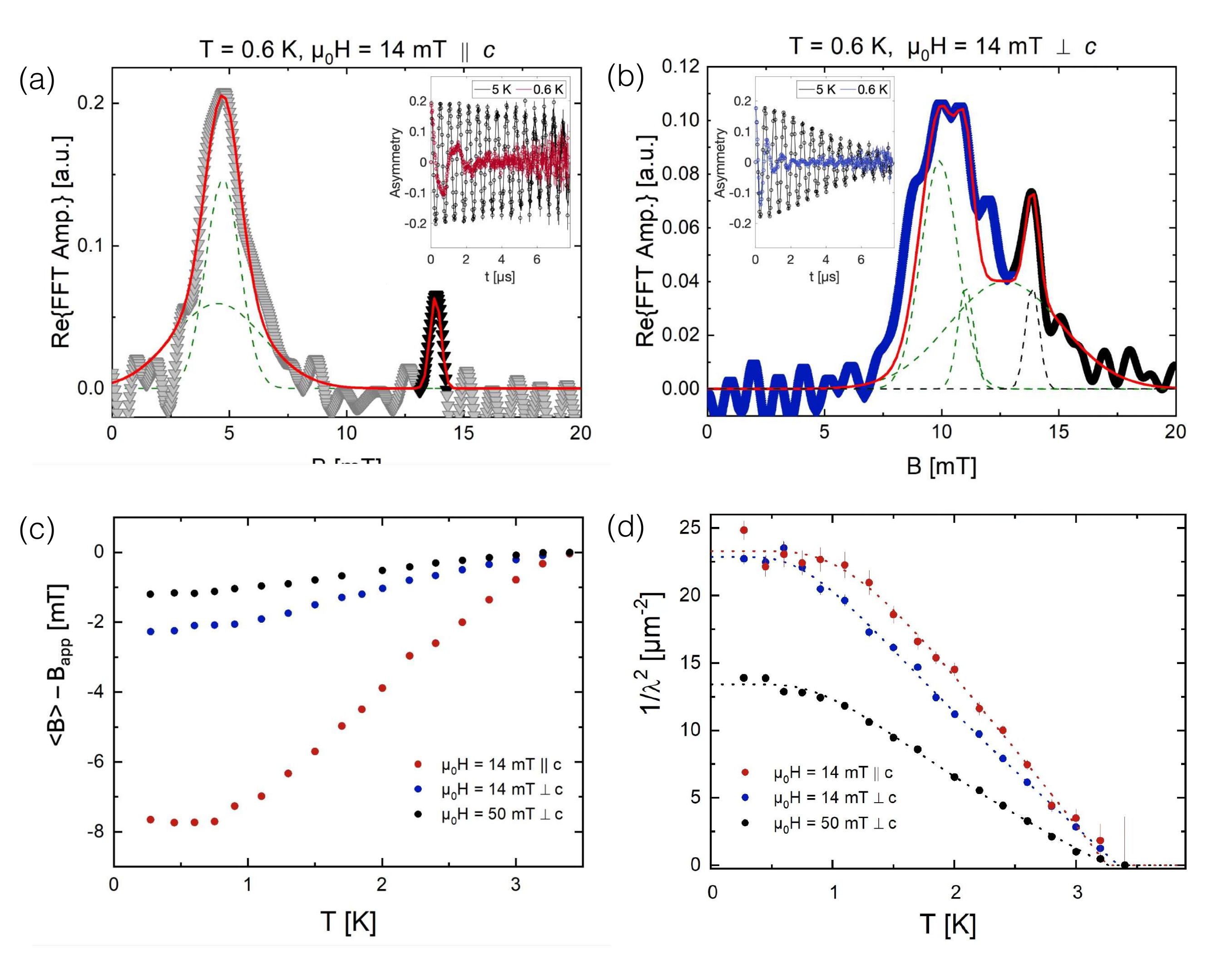}
\caption{\label{fig:TF_muSR}Fast Fourier transformation (FFT) of the
\tfmu\ time-dependent asymmetry below $T_c$ in a 14-mT field applied
parallel- (a) and perpendicular (b) to the $c$-axis at ambient pressure.
Representative asymmetries below and above $T_{c}$ are reported in the
insets. In both cases, the increased relaxation rate below $T_c$ reflects
the field modulation due to the superconducting vortices. Note the much
lower relaxation rate for $T > T_c$ in the $H^{\parallel}$ case. The red
lines represent the frequency-domain response of a model using the
best-fit parameters of the time-dependent asymmetry. The dashed green-
and black lines represent the superconducting- and normal state components,
respectively. In the perpendicular configuration a further frequency
component was required to resolve the tiny peak splitting (see text).
Note the substantial diamagnetic shift in panel (a), attributed to an
almost null pinning force in the $H^{\parallel}$ case. Diamagnetic field
shift $\langle B \rangle - B_\mathrm{app}$ (c), with
$B_\mathrm{app}$ being the applied field intensity, and $1/\lambda^{2}$ (d) vs.\  temperature for a magnetic field applied parallel- and perpendicular to
the $c$-axis, as derived by assuming a conventional AVL. Continuous lines
represent $s$-wave fitting models (see text).}
\end{figure*}

\noindent\textbf{\textcolor{black}{\musr\ at ambient pressure}}\\[0.3ex]
To investigate the superconducting ground-state properties of \vga,
we carried out systematic tempera\-ture\--de\-pen\-dent \tfmu\ measurements.
A magnetic field of 14\,mT, i.e., well above the $H_{c1}$ value reported
in the literature~\cite{Cruce1974,Cheng2024} and confirmed by our measurements
at 2\,K (see Fig.~\ref{fig:Hc1}), was applied parallel- and orthogonal
to the $c$-axis by using a FC protocol. In the orthogonal configuration, we also applied a magnetic field of 50\,mT. Representative \tfmu\ spectra collected in the normal- and superconducting states of \vga\ are shown in the insets of Fig.~\ref{fig:TF_muSR}a and Fig.~\ref{fig:TF_muSR}b. In the superconducting state (i.e., $T<T_{c}$), the
development of a flux-line lattice (FLL) causes an inhomogeneous field distribution and, thus, gives rise to an additional damping in the \tfmu\ spectra~\cite{Yaouanc2011}. In both the parallel and perpendicular configurations the time-de\-pen\-dent asymmetry was modelled by the following equation~\cite{Maisuradze2009}:
\begin{equation}
    \label{eq:TF_muSR}
    A_\mathrm{TF}(t) = \sum\limits_{i=1}^n A_i \cos(\gamma_{\mu} B_i t + \phi) e^{- \sigma_i^2 t^2/2} +
    A_\mathrm{bg} \cos(\gamma_{\mu} B_\mathrm{bg} t + \phi).
\end{equation}
Here $A_{i}$, $A_\mathrm{bg}$ and $B_{i}$, $B_\mathrm{bg}$ 
are the initial asymmetries and local fields sensed by the muons
probing the vortex state and the non superconducting part of the sample
(or parts of the sample holder), respectively, $\gamma_{\mu}$/2$\pi$ = 135.53\,MHz/T 
is the muon gyromagnetic ratio, and $\phi$ is a shared initial phase.
Here, $\sigma_{i}$ represents the Gaussian relaxation rate of the $i$th
component. This is temperature-independent in the normal state but,
below $T_c$, it starts to increase due to the onset of the FLL (see
insets in Fig.~\ref{fig:TF_muSR}a and Fig.~\ref{fig:TF_muSR}b). We recall that the fast
Fourier transform (FFT) of the time-dependent asymmetry represents the
probability distribution density of the local field $p(B)$. In general,
in the superconducting state, the $p(B)$ distribution is material
dependent. In case of a symmetric $p(B)$, one oscillation (i.e., $n = 1$)
is sufficient to describe the \tfmu\ spectra.
However, for an asymmetric $p(B)$, two or more oscillations (i.e., $n \ge 2$)
are required. Here, we find that Eq.~\eqref{eq:TF_muSR} with $n = 2$
describes the experimental data quite well (see green dashed lines in
the main panels of Fig.~\ref{fig:TF_muSR}a and Fig.~\ref{fig:TF_muSR}b).
We evidence here four notable aspects:

%
\begin{figure*}[t]
	\centering
	\vspace{-1ex}%
	\centering
\raisebox{0.11\height}{\includegraphics[width=0.33\linewidth]{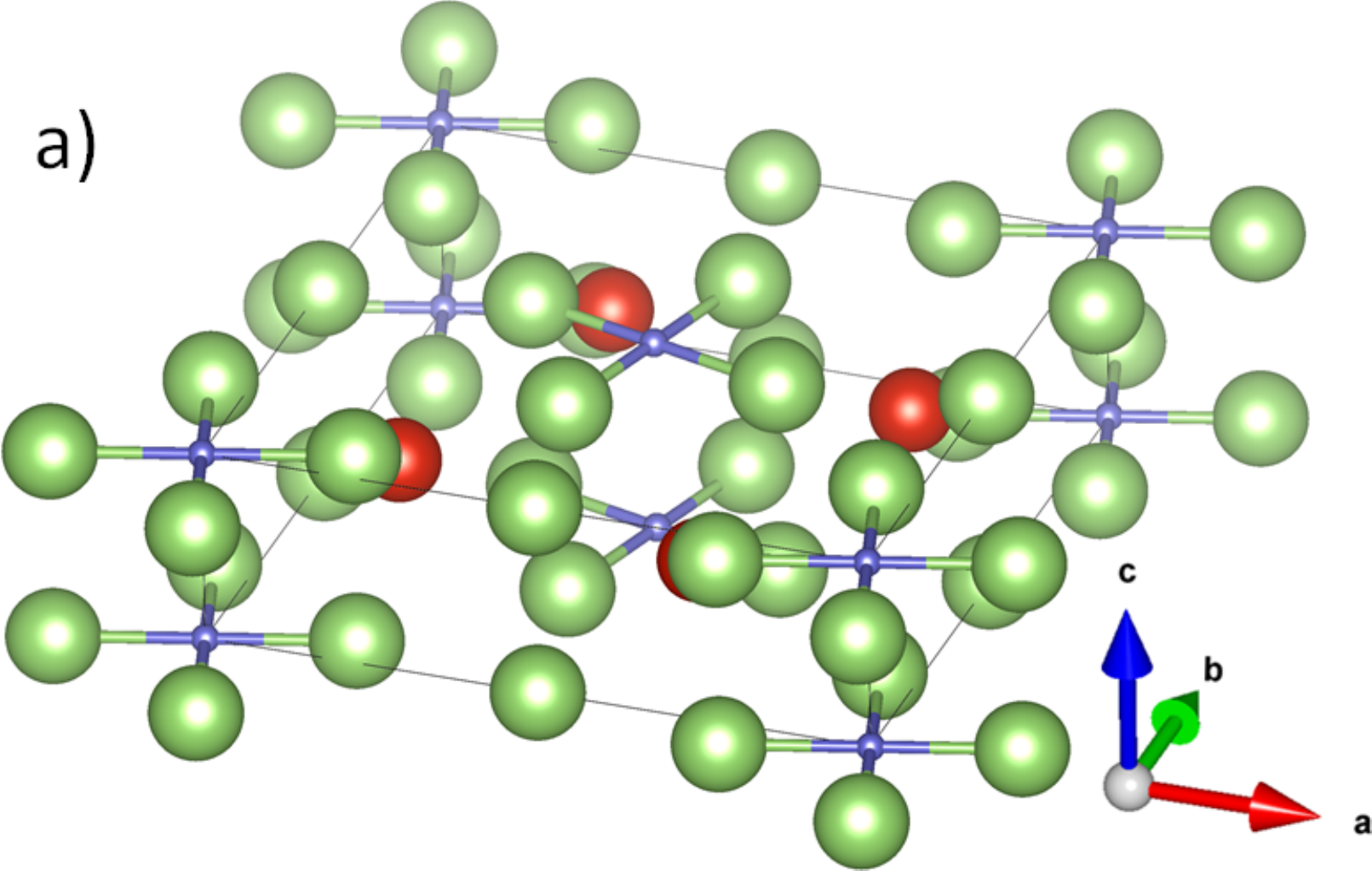}}\hspace{4mm}
\includegraphics[width=0.29\linewidth]{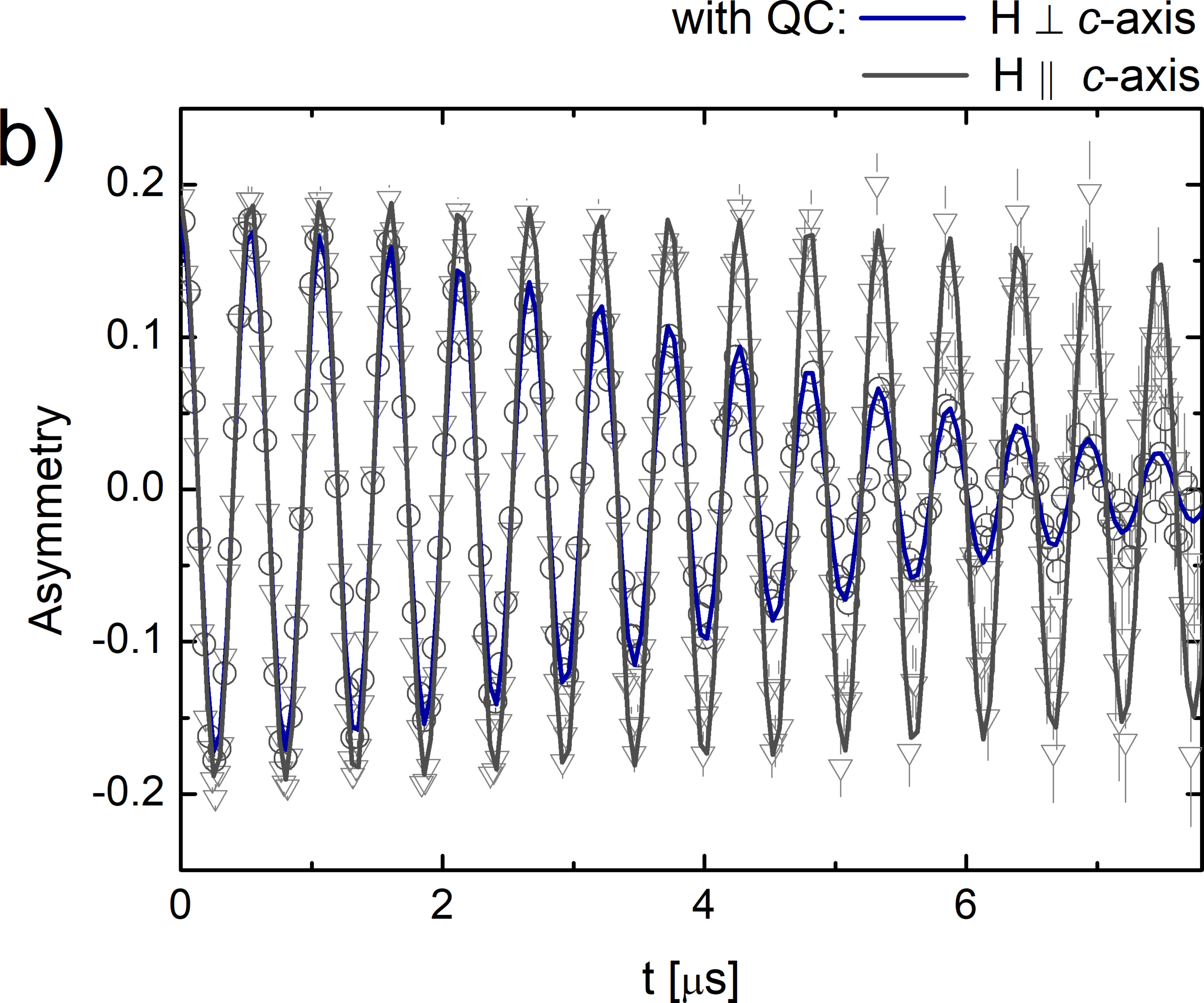}\hspace{4mm}
\includegraphics[width=0.29\linewidth]{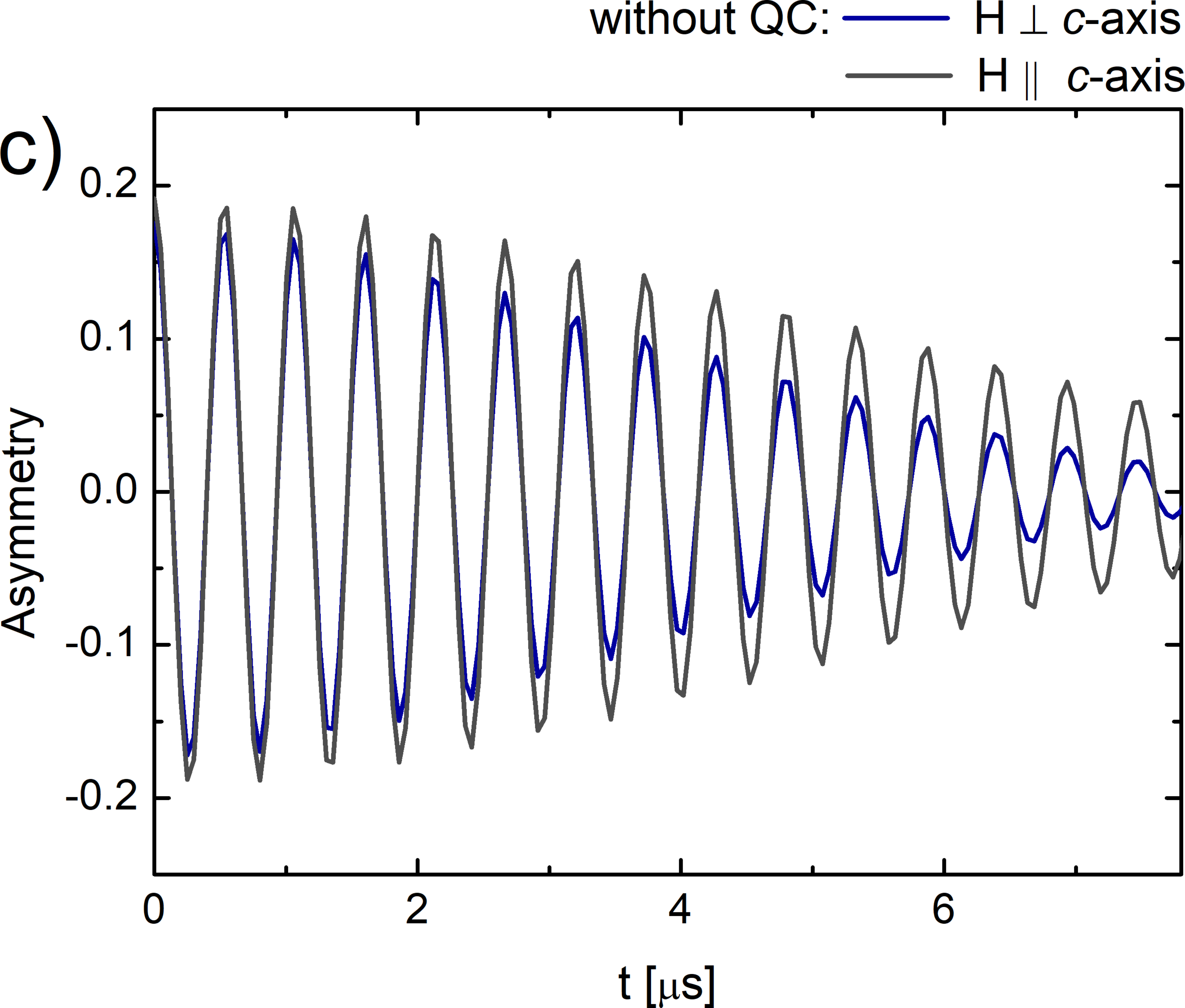}
	\caption{\label{muons} DFT$+\mu$ prediction of the most energetically
	favorable muon implantation site, which turns out to be fully anisotropic (a).
	Panel (b) shows the simulated asymmetries including the
	quadrupolar contribution (QC) vs the measured asymmetries at $T = 5$\,K (i.e., above $T_{c}$)
	for an applied transverse field $\mu_0H^{\parallel} = 14$\,mT ($\triangledown$) and
	$\mu_0H^{\perp} = 14$\,mT ($\circ$). Panel (c) shows the simulated
	asymmetries without QC for a comparison.}
\end{figure*}
%

Firstly, the muon precession is strongly damped in the superconducting state, suggesting low values of the magnetic penetration depth (see below), thus confirming the very high quality of our samples.

Secondly, in both the superconducting- and normal states, the ti\-me\--de\-pen\-dent asymmetry is very different in the parallel vs.\ the perpendicular configuration (see insets in Fig.~\ref{fig:TF_muSR}). The superconducting state is discussed in detail in the following sections. As to the normal state, we remark that the muon depolarization in the $H^{\perp}$ configuration is much larger than that measured in the $H^{\parallel}$ case (see also the NMR section above). Recently, Bonf\`{a} \textit{et al.}~\cite{Bonfa2022} have clarified the  perturbation produced by implanted muons on the lattice and the EFG at the neighboring sites. Such perturbation is more effective when the muon implantation site is close to atoms whose nuclei have non-zero quadrupole moments, as is here the case for Ga and V. Following this proposal, density functional theory (DFT) was used to calculate the most likely implantation sites for positive muons, here modeled as hydrogen impurities (DFT$+\mu$ procedure~\cite{Blundell2023,Ifeanyi2025}), and the (perturbed) EFG at the various nuclei. The nuclear contribution to the muon spin relaxation is finally obtained with the code developed in Ref. \onlinecite{BonfaUNDI}. In Fig.~\ref{muons}a we show the result of such calculation: the predicted muon implantation sites lie on the Ga planes, where the implanted muons are bounded with four Ga first-neighbour atoms, whereas the out-of-plane second and third neighbours of the muon, respectively Ga and V nuclei, are more than 3~\AA{} apart, resulting in a less effective dipolar coupling. This scenario confirms that the coupling of the implanted
muons to the environment is essentially planar, i.e. \emph{fully anisotropic}. In Fig.~\ref{muons}b-c we show the calculated muon asymmetries~\footnote{The calculation was carried out following the approach proposed by Celio~\cite{Celio1986}, considering a Hilbert space with 4 first- and 8 second nearest-neighbor Ga nuclei.}, with and without the quadrupolar contribution (QC) of the Ga nuclei, for both the $H^{\perp}$ and $H^{\parallel}$ configurations. It is worth noting that (i) the quantum entanglement with the dipolar moments of Ga nuclei is the main responsible for the large depolarization rate. (ii) The largest component of the EFG tensor at Ga nuclei is the one in the \textit{ab}-plane, which makes the muon perturbation of the EFG less effective~\cite{Bonfa2022}. Thus, for $H^{\perp}$, the pure nuclear dipolar contribution can be recovered, while for $H^{\parallel}$ the anisotropy of the EFG tensor contributes significantly to reduce the muon-spin depolarization. The excellent agreement between the simulated asymmetries
in the presence of QC and the corresponding experimental data sets
(Fig.~\ref{muons}b) confirms the picture of a very large and
anisotropic EFG at the Ga nuclei. As an additional check on the accuracy
of our numerical calculations, we also calculated the EFG and quadrupolar
coupling constants at the V nuclei. The calculated value of $2\nu_Q = 214 \pm 2$\,kHz
is close to the experimental value of $2\nu_Q = 274\pm 2$\,kHz (see Fig.~\ref{fig:NMR_results}a).
Hence, our DFT model is consistent with both the \musr\ and NMR data. 
%

Thirdly, the corresponding fast Fourier transforms show a main superconducting
peak and a minor peak at the applied field value. The latter is ascribed
to a tiny non superconducting part of the sample (or parts of the sample holder). The resulting low-temperature lower-limit of the superconducting volume fraction
is $V_\mathrm{sc}(\mathrm{0.3\,K}) =1- A_\mathrm{bg}/A_\mathrm{tot} \simeq 95(1)$\%,
in agreement with the superconducting shielding fraction estimated by
dc magnetization, thus further confirming the very high quality of the
investigated samples.

Fourthly, in the parallel-field configuration, the normal peak is $\sim 10 $\,mT
away from the superconducting one, thus indicating an almost null pinning force in this case.
Conversely, in the perpendicular-field configuration, the shift is small, 
indicative of strong pinning. In Fig.~\ref{fig:TF_muSR}c the temperature evolution of the superconducting field shift in both field configurations is represented. 
Further, for $\mu_0H^{\perp} = 14$\, mT we observe also a tiny peak splitting in the superconducting component (see main panel of Fig.~\ref{fig:TF_muSR}b), while it is completely absent for $\mu_0H^{\perp} = 50$\, mT (not shown). We calculate the resulting depolarization rate $\sigma$ from the second moment of the internal
field distribution $P(B)$ by assuming a standard Abrikosov vortex lattice
(AVL) in both configurations. The resulting effective Gaussian relaxation
rate can be written as~\cite{Maisuradze2009}:
\begin{gather}
\frac{\sigma^2}{\gamma_\mu^2} = \sum_{i=1}^2 A_i [\sigma_i^2/\gamma_{\mu}^2 - \left(B_i - \langle B \rangle\right)^2]/(A_1 + A_2),\\
\text{where} \quad \langle B \rangle = (A_1\,B_1 + A_2\,B_2)/(A_1 + A_2). \nonumber
\end{gather}

By considering that, in the 0--5\,K range investigated here, the
nuclear relaxation rate $\sigma_\mathrm{n}$ is essentially independent
of temperature, the Gaussian relaxation rate in the SC state can be
extracted from the total depolarization rate $\sigma$ via a quadrature subtraction:
\begin{equation}
\sigma_\mathrm{sc} = \sqrt{\sigma^{2} - \sigma^{2}_\mathrm{n}}.
\end{equation}
In \vga, the upper critical fields measured in samples from the same origin $\mu_0H_\mathrm{c2}^{\parallel}(0) \sim 0.49$\,T 
and $\mu_0H_\mathrm{c2}^{\perp}(0) \sim 0.57$\,T ~\cite{Xiaofeng2024} are significantly larger than the applied TF field (14 and 50\,mT respectively). Hence, we can ignore the effects of the overlapping vortex cores when extracting the magnetic penetration depth from the measured $\sigma_\mathrm{sc}$. The effective magnetic penetration depth $\lambda_\mathrm{eff}(T)$ can then be calculated by using the well-known relation~\cite{Barford1988,Brandt2003}:
\begin{equation}
\label{eq:brandt}
\frac{\sigma_\mathrm{sc}^2(T)}{\gamma^2_{\mu}} = 0.00371\frac{\Phi_0^2}{\lambda_\mathrm{eff}^4(T)},
\end{equation}
where $\Phi_0 = 2.07 \times 10^{3}$\,T\,nm$^{2}$ is the magnetic flux quantum. \\

\begin{figure*}[t]
	\centering
	\vspace{-1ex}%
	\centering
\includegraphics[width=0.9\linewidth]{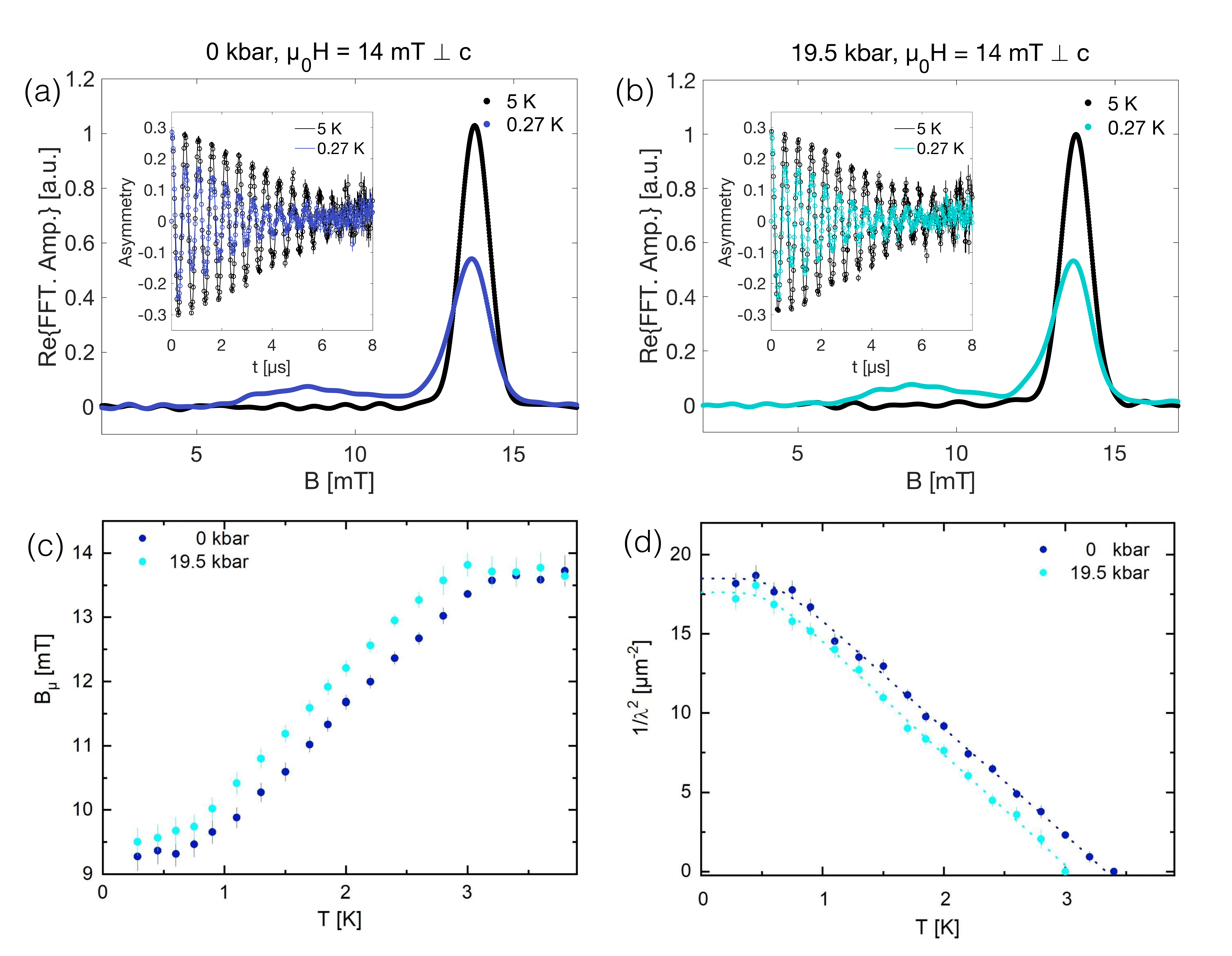}
	\caption{\label{fig:GPD_muSR}\tfmu\ relaxation above and
	below $T_c$ in a 14-mT field applied perpendicular to the $c$-axis
	at zero- (a) and maximum pressure, 19.5\,kbar (b). The time-domain
	spectra above and below $T_c$ for both pressures are given in the insets. 
	In both cases, the \musr\ signal is dominated by the pressure-cell
	contribution, here appearing as an enhanced relaxation of asymmetry
	(compare with insets in Fig.~\ref{fig:TF_muSR}). This is reflected also
	in the respective FFT spectra (main panels), with the main peak due
	to the pressure cell and the broad hump at the left representing
	the sample contribution. Diamagnetic field shift and $1/\lambda^{2}$ (d) vs.\ temperature measured at zero- and at 
	maximum pressure (19.5\,kbar). Dashed lines in panel (b)
	represent a fit with a single $s$-wave SC pairing model.
	The applied pressure reduces slightly $T_c$, but it does not change
	the nature of the pairing nor its parameters.}
\end{figure*}

\begin{figure}[!thp]
	\centering
	\includegraphics[width=0.9\linewidth]{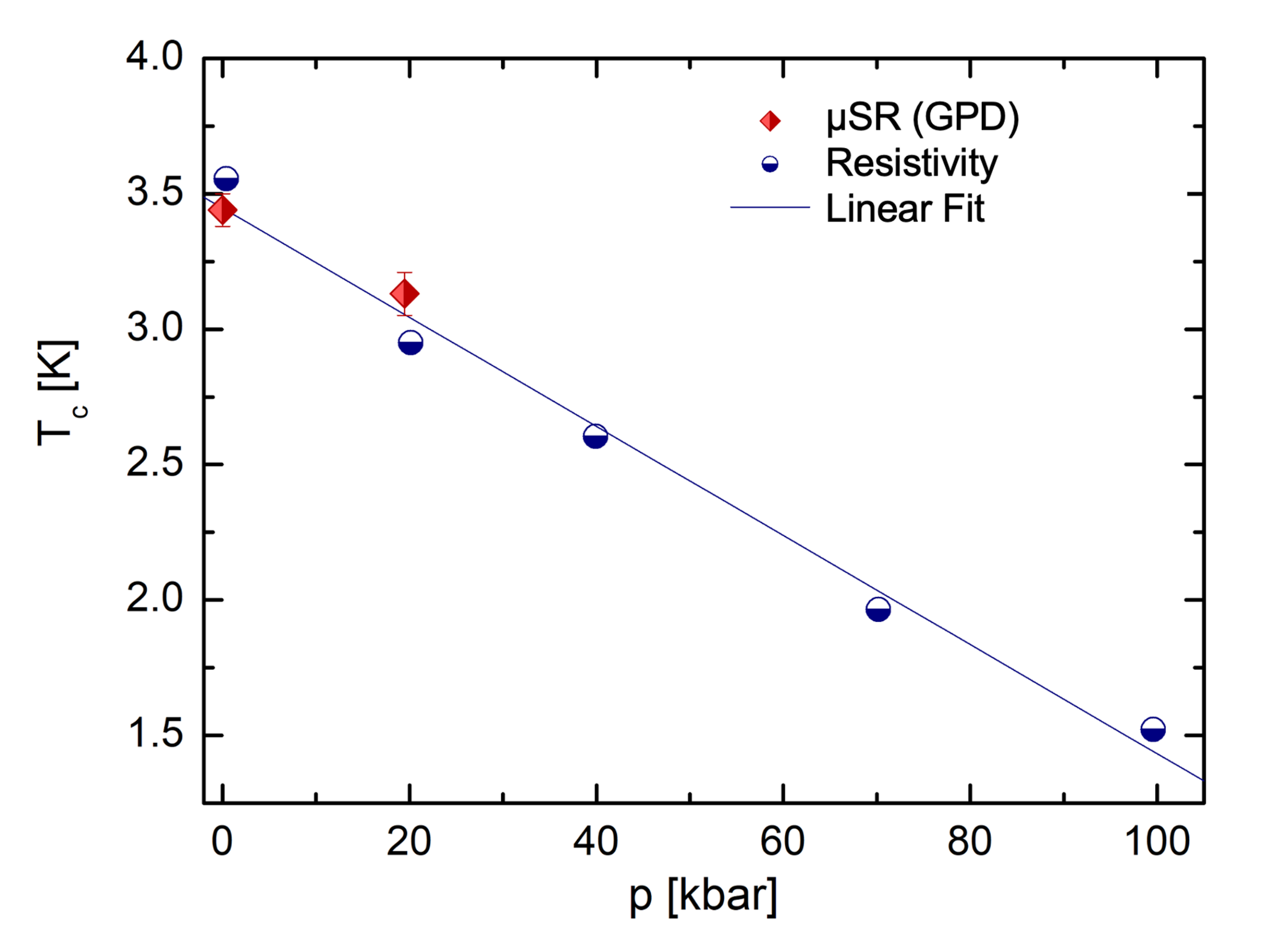}
	\caption{\label{fig:Tc_vs_press}Critical superconducting temperature
	of \vga\ vs. applied pressure. Resistivity~\cite{Huang2025}
	and {\textmu}SR results (this work) suggest a linear decrease of
	$T_c$ with pressure, with an almost identical slope.}
\end{figure}

\noindent\textbf{\textcolor{black}{\label{ssec:musr-vsp}\musr\ under applied hydrostatic pressure}}\\[0.3ex]%
To gain further insight into the superconducting properties of \vga, 
we performed \tfmu\ measurements in an applied pressure up to 19.5\,kbar.
In Fig.~\ref{fig:GPD_muSR}a and Fig.~\ref{fig:GPD_muSR}b we present
the time-dependent asymmetry and its fast Fourier transform at zero- and
at maximum pressure, both below and above $T_c$. The pressure cell affects
strongly the FFT profile by significantly broadening the superconducting
peak compared with the data shown in Fig.~\ref{fig:TF_muSR}a and
Fig.~\ref{fig:TF_muSR}b. For this reason, the optimal fit of the
superconducting peak was obtained by assuming a skewed Gaussian (skG)
field distribution~\cite{SKW2016}. In this case, the time-dependent
asymmetry was fitted by the following equation:
\begin{equation}
\begin{split}
	\label{eq:TF_SKG}
	A_\mathrm{TF}(t) &= A_\mathrm{0} \bigl[ \left(1-f_\mathrm{pc} \right) \cdot \mathrm{skG}(t)\bigr. +\\
	                 &+ \bigl. f_\mathrm{pc} \cos(\gamma_{\mu} B_\mathrm{pc} t + \phi)e^{- \sigma_\mathrm{pc}^2 t^2/2} e^{- \Lambda_\mathrm{bg} t} \bigr].
\end{split}
\end{equation}
Here, $\mathrm{skG}(t)$ is the frequency-to-time  
Fourier transform of the skewed Gaussian field distribution,
while $A_\mathrm{pc}$, $B_\mathrm{pc}$ and $\sigma_\mathrm{pc}$
refer to the contribution of the muons implanted in the pressure cell
($A_\mathrm{pc} / A_\mathrm{0} \sim$ 70\%). $\sigma_\mathrm{bg}$ was kept fixed at 0.26\,{\textmu}s$^{-1}$, its known temperature-independent value in the 0.25--20\,K range~\cite{ShermadiniHPR}.
The additional simple exponential relaxation term $e^{-\Lambda_\mathrm{bg} t}$ accounts for the fluctuating dipole moments of electronic origin. Since such relaxation
increases slightly below 1\,K, we had to fix $\Lambda_\mathrm{bg}$ to its previously measured value at each of the corresponding temperatures,
as reported in Ref.~\onlinecite{ShermadiniHPR}. Due to the overwhelming
weight of the cell background, the comparison of Dolly- (no background) with the GPD results at zero pressure was essential to estimate the contribution of the pressure cell to the \musr\ signal. In Fig.~\ref{fig:GPD_muSR}c and Fig.~\ref{fig:GPD_muSR}d we show the temperature evolution of the
diamagnetic field shift and of $\lambda^{-2}$, extracted from the
superconducting depolarization rate by means of Eq.~(\ref{eq:brandt}). \\

\noindent\textbf{\textcolor{black}{\label{ssec:musr-ns}Superfluid density}}\\[0.3ex]
Figures~\ref{fig:TF_muSR}d and \ref{fig:GPD_muSR}d summarize the
temperature-de\-pen\-dent inverse-square of the magnetic penetration depth,
which is proportional to the superfluid density, i.e.,
$\lambda_\mathrm{eff}^{-2}(T) \propto \rho_\mathrm{sc}(T)$. In the clean
limit and for an $s$-wave pairing symmetry, this is described
by the following equation:
\begin{equation}\label{eq:ns}
\frac{\lambda^{2}(0)}{\lambda^{2}(T)}= \left[ 1-2 \int_{\Delta (0)} ^{\infty} \left( - \frac{\partial f }{ \partial E} \right) \cdot \frac{E \, \mathrm{d}E}{\sqrt{E^{2}-\Delta^{2}(T)}} \right],
\end{equation}
where $f = (1+e^{E/k_\mathrm{B}T})^{-1}$ is the Fermi function and
$\Delta(T) = \Delta_0 \mathrm{tanh} \left[ 1.82[1.018(T_\mathrm{c}/T-1)]^{0.51} \right]$~\cite{Tinkham1996,Carrington2003}, with $\Delta_0$ being the SC gap value at 0\,K.

As shown in Figs.~\ref{fig:TF_muSR}d and ~\ref{fig:GPD_muSR}d, the constant
$1/\lambda^{2}(T)$ values at low-$T$ clearly suggest a fully-gapped superconducting state. More interestingly, around $T_{c}/2$ an inflection point appears for
$\mu_0H^{\perp} = 14$\,mT in both GPD-data sets. Such an inflection point is expected to appear in superconductors with two weakly coupled superconducting bands,
as for instance is the case of MgB$_2$~\cite{Golubov02}.
Similarly, we analyzed our datasets by assuming the superfluid
density to be due to a linear combination of two independent
contributions from bands with a negligible interband scattering~\cite{Golubov02,Zurab15}:
\begin{equation}\label{eq:ns2gap}
\frac{\lambda^{2}(0)}{\lambda^{2}(T)}= \omega \cdot \frac{\lambda_{1}^{2}(0, \Delta_{0,1})}{\lambda_{1}^{2}(T,\Delta_{0,1})} + (1-\omega) \cdot \frac{\lambda_{2}^{2}(0, \Delta_{0,2})}{\lambda_{2}^{2}(T,\Delta_{0,2})},
\end{equation}
where $\lambda(0)$ is the zero-temperature limit of the penetration
depth, $\Delta_{0,i}$ is the value of the $i$-th superconducting gap
($i = 1$, 2) at $T = 0$\,K and $\omega$ is a weighting factor,
representing the relative contribution of each band.
All the fitting parameters are summarized
in Table~\ref{tab:lambda}. It is worth noting that, for these three
datasets, a global fit was performed by assuming a negligible variation
of the relative weight of the two bands up to 20\,kbar.
The resulting global parameter $\omega = 0.44(8) $ suggests an
almost equal contribution of the two bands. 
We further note that, in the zero-temperature limit, both the superconducting
gap and the magnetic penetration depth are almost insensitive to the
applied pressure (up to $\sim 19.5$\,kbar). Conversely, $T_{c}$
decreases slightly with increasing pressure, with a slope of $\mathrm{d}P/\mathrm{d}T =-0.020(1)$\,kbar/K.
Such decrease in $T_c$ is in good agreement with the resistivity data
measured up to 100\,kbar (see Fig.~\ref{fig:Tc_vs_press}) on samples
from the same batch~\cite{Huang2025} and/or reported in a recent work~\cite{Cheng2024}. 

We recall that for most superconductors, $T_{c}$ generally decreases
with pressure~\cite{Drickamer1965,Shiroka2022}. Indeed, one can write 
$T_{c} \sim \sqrt{k/M} \cdot \text{e}^{\frac{-k}{\eta}}$, 
%
where $k$ is a spring constant (a lattice term), $\eta$ is a purely electronic term,
and $M$ is the atom mass. Generally, the pressure-induced lattice stiffening $k$ dominates
over small changes to the electronic term $\eta$. When pressure is applied, the spring constant $k$ increases.  However, the increase in the prefactor is outweighed by the decrease in the exponential. The overall effect is a decrease in $T_c$.

Interestingly, no inflection points could be detected for
$\mu_0H^{\parallel} = 14$\,mT and $\mu_0H^{\perp} = 50$\,mT.
In both cases, a single $s$-wave gap model in the clean limit
could satisfactorily fit the experimental data. 
In the 50-mT case, depending on the geometry of the circular motions on the Fermi surface, one of the bands may dominate the SC quasiparticle excitations, thus resulting in the vanishing of the smaller gap.
As to the 14-mT case, the vanishing of the small gap could be the
fingerprint of a $k$-space gap anisotropy due to the tetragonal
structure of \vga.\\

\begin{table*}[tbh]
\centering
\renewcommand{\arraystretch}{1.2}
\caption{\label{tab:lambda} Fit parameters $\lambda(0)$ and $\Delta(0)$ at 0\,K,  
as resulting from a fit to Eq.~(\ref{eq:ns}) and/or~(\ref{eq:ns2gap}) as shown in Figs.~\ref{fig:TF_muSR}d and~\ref{fig:GPD_muSR}d. The high-pressure (GPD) datasets,
as well as the data for $\mu_0H^{\perp} = 14$\,mT were analyzed by means
of a global two $s$-wave model, with the relative weight $\omega = 0.44(8)$ being a
global fit parameter (see text for details).}
\begin{ruledtabular}
 	\begin{tabular}{ lcccl }
	$p$ [kbar] & $\lambda(0)$ [nm] & $T_{c}$\,[K] & $\Delta(0)$ [meV]  \\ \midrule
	0 & $\lambda^{\parallel}$ (14\,mT) \hfill 207(2) & 3.26(5) &  $\Delta^{\parallel}$(14\,mT) \hfill 0.43(1)  &\\ 
	0 & $\lambda^{\perp}$\,(14\,mT) \quad \hfill 209(1)  & 3.35(2) & $\Delta^{\perp}_{1}$\,(14\,mT) \quad\hfill 0.27(4)  & \\
	&        &         & $\Delta^{\perp}_{2}$\,(14\,mT) \quad\hfill 0.45(4)  &\\
	0 & $\lambda^{\perp}$ (50\,mT)  \quad \hfill 273(1)  & 3.25(2) & $\Delta^{\perp}$(50\,mT) \hfill 0.36(4)&\\
	\midrule
	0~~~~~~(in press.\ cell) & $\lambda^{\perp}$\,(14\,mT) \quad \hfill 233(3)  & 3.34(5) & $\Delta^{\perp}_{1}$\,(14\,mT) \quad\hfill 0.23(5)  & \\
	&        &         & $\Delta^{\perp}_{2}$\,(14\,mT) \quad\hfill 0.48(7)  &\\
	19.5 (in press.\ cell) & $\lambda^{\perp}$\,(14\,mT) \quad \hfill 238(3)   & 3.06(7)  & $\Delta^{\perp}_{1}$\,(14\,mT) \quad\hfill 0.21(5)  &\\
	&         &          & $\Delta^{\perp}_{2}$\,(14\,mT) \quad\hfill 0.44(7)  &\\
	\end{tabular}
\end{ruledtabular}
\end{table*}
\vspace{15pt}
\noindent{\large\textbf{Discussion}}\\[0.3ex]
\noindent\textbf{\textcolor{black}{Search for topological signatures in the normal state }}\\[0.3ex]
It is of interest to check if the topological properties of \vga\ give
rise to observable experimental signatures in its normal phase.
We first give a brief overview of the existing theory. It is known that
topological systems generally exhibit a characteristic diamagnetic
response~\cite{koshino2016magnetic}. In particular, at high temperatures,
the NMR shift arising from a single (Weyl or Dirac) point
node is predicted to follow a logarithmic diamagnetic temperature
dependence~\cite{okvatovity2016anomalous}:
\begin{equation*}
K(\mu(T), T) \approx \frac{\mu_{0} e}{4 \pi^{2} \hbar}\left[\frac{g \mu_\mathrm{B}}{\hbar v_\mathrm{F}} \mu - \frac{e v_\mathrm{F}}{3} \ln \left(\frac{W}{\max \left[|\mu(T)|, k_\mathrm{B} T\right]}\right)\right]
\label{eqn:knightshiftmasterequation}
\end{equation*}
where $W\gg\mu$ denotes a sharp high-energy cutoff regularizing the theory,
$\mu_{0}$ is the chemical potential at $T=0$, $\mu(T)$ is the
temperature-dependent chemical potential away from the point node,
$v_\mathrm{F}$ is the Fermi velocity, and $\hbar$, $e$, $g$,
$\mu_\mathrm{B}$ are the usual physical constants. Likewise, at high
temperatures, $1/T_{1}T$ is predicted to follow a power-law
dependence given by~\cite{okvatovity2019nuclear}:
\begin{equation*}
\begin{aligned}
    \frac{\hbar}{k_{\mathrm{B}} T_{1} T}&= \frac{52.7k_{\mathrm{B}}  \pi \mu_{0}^{2} \gamma_{\text{n}}^{2} e^{2}}{(2 \pi)^{6} \hbar v_{\mathrm{F}}^{2}}\\
&\times\begin{cases}\left(\dfrac{k_{\mathrm{B}} T}{\hbar}\right)^{2} \dfrac{\pi^{2}}{3} \ln \left(\dfrac{4 k_{\mathrm{B}} T}{\hbar \omega_{0}}\right), & \mu \ll k_{\mathrm{B}} T \\
    \left(\dfrac{\mu}{\hbar}\right)^{2} \ln \left(\dfrac{2 \mu}{\hbar \omega_{0}}\right), & \mu \gg k_{\mathrm{B}}T\end{cases}
\end{aligned}
\label{eqn:relaxationmasterequation}
\end{equation*}

In case of Kramers nodal-lines semimetals, to our knowledge, a
comprehensive theory of the expected NMR behavior has not yet been
developed. However, previous measurements on the nodal-line semimetal
ZrSiTe~\cite{ZrSiTe2021tian}, which exhibits a single nodal line,
reveal a characteristic ``V-shaped'' feature, with minima in both the
NMR shift and relaxation rate occurring at a temperature $T_\mathrm{min}$,
where the temperature-dependent chemical potential
$\mu(T) \propto |T - T_\mathrm{min}|$ goes to 0. 

Now, turning to the case of \vga, our measured $1/T_{1}T$ is practically
temperature-independent (see Fig.~\ref{fig:NMR_results}c). Hence, the
$1/T_{1}T$ results do not exhibit any of the experimental signatures
expected for a single point-node or a nodal line. The only hint of
topological properties occurs in the slight logarithmic increase of
the NMR Knight shift above 90\,K, which corresponds to a chemical potential
$\mu = k_\mathrm{B}T \approx 7.76$\,meV. However, from previous
calculations, the closest lying point node is at $\sim 70$\,meV,
hence casting doubts on the topological origin of this feature. Furthermore,
the increase in $K$ is much smaller than the NMR line width and there
is no accompanying change in $1/T_{1}T$. Consequently, the increase in
NMR shift above 90\,K is most likely due to small changes in the
quadrupole interaction caused by tiny temperature-induced changes in
the crystal structure. \\

\begin{figure}[t]
	\centering{}
	\vspace{-2mm}
        \includegraphics[width=0.9\linewidth]{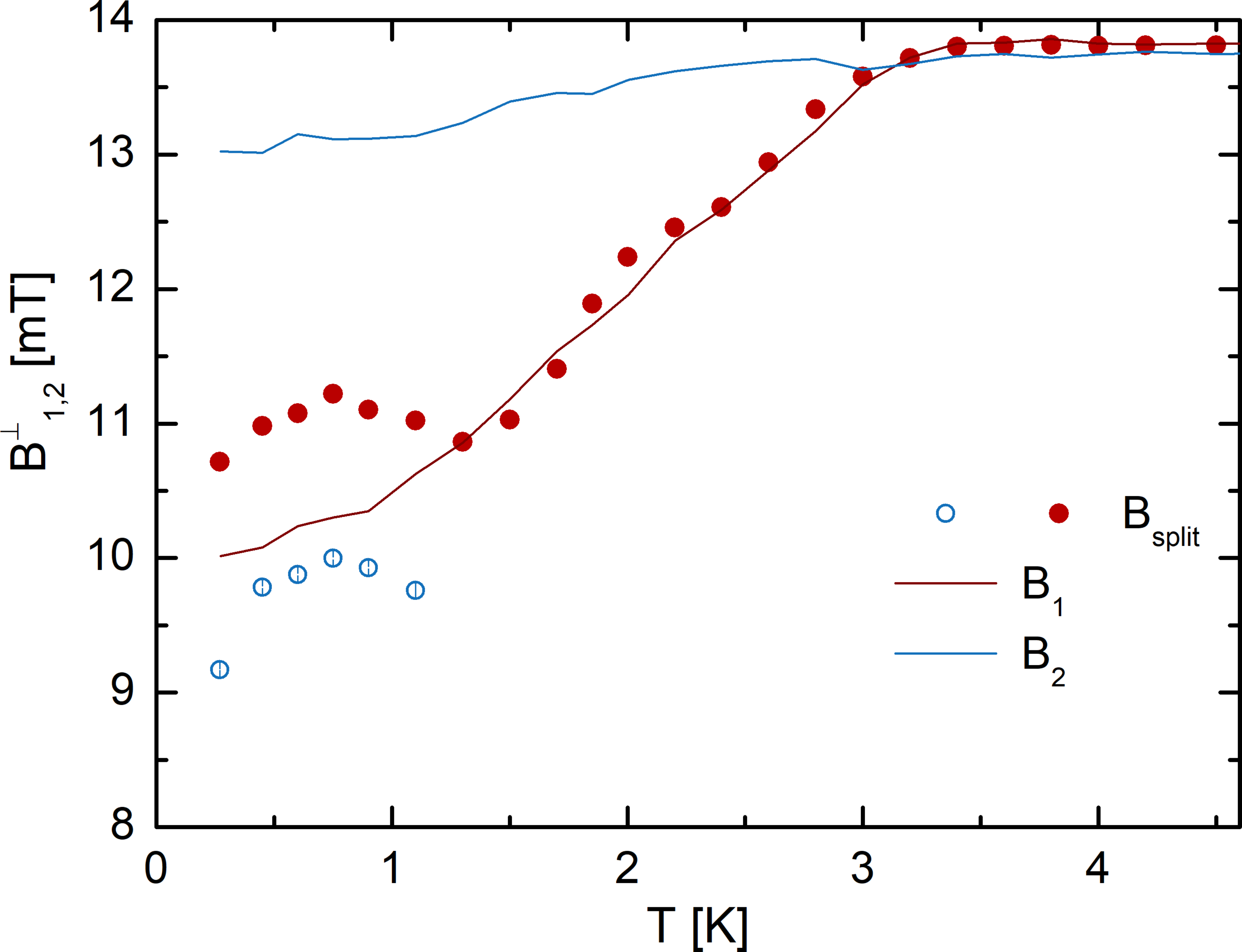}
	\caption{\label{fig:Bsplit} Temperature dependence of internal fields $B_{1}$ and $B_{2}$. The continuous lines represent the temperature evolution of $B_{1}$ and $B_{2}$ obtained by fitting the time dependent asymmetry for the $\mu_0H^{\perp}=14 $ mT field configuration through Eq.~\eqref{eq:TF_muSR}. Blue and red dots represent the temperature evolution of the peak splitting obtained by directly fitting the FFT of the time dependent asymmetry.}
\end{figure}

\noindent\textbf{Anisotropic superconductivity in \vga}\\[0.3ex]
\vga\ has a tetragonal structure with the $c$-axis being less
than one third of the in-plane unit cell parameter (2.6831\,\AA\ vs.\
8.936\,\AA~\cite{Lobring2002}). This pronounced difference in the
lattice parameters is most likely responsible for the highly anisotropic
electronic properties we observe in the normal- and superconducting states.
Yet, \tfmu\ measurements in the superconducting state reveal a more
complex picture.

Firstly, we consider the anisotropy in the superconducting state, as
determined from the magnetic penetration depth and superfluid density.
In the $T\rightarrow 0$\,K limit, the magnetic penetration depth and,
consequently, the superfluid density both show an almost negligible
anisotropy [see Table~\ref{tab:lambda}].
On the other hand, the temperature dependence of the superfluid density
shows clear differences between the in-plane $H^{\parallel}$
and out-of-plane $H^{\perp}$ configurations (Fig.~\ref{fig:TF_muSR}d).
These different behaviors may explain why in the $H^{\parallel}$
case a single gap is found, while for $H^{\perp}$ a double gap provides
the best fit. Angle resolved photoemission spectroscopy (ARPES)
measurements evidence two discernible bands with an almost negligible
dispersion along the $k_{x}$ and $k_{y}$ directions cross the Fermi
level close to the $Z$ and $\Gamma$ points~\cite{Cheng2024}. Therefore,
in an applied magnetic field orthogonal to the $c$-axis ($H^{\perp}$),
the induced screening currents in the \textit{ac}-plane are likely
sensitive to both gaps. On the other hand, this might not be the case
when the field is parallel to the $c$-axis and the induced currents lie
in the $ab$-plane.

Secondly, the anisotropy of superconductivity in \vga\ can
be observed from the anisotropy of pinning forces. The unexpectedly
large diamagnetic field shift in the $H^{\parallel}$ configuration
suggests a null pinning force. On the other hand, the 
small diamagnetic field shift in the $H^{\perp}$ configuration suggests a non-negligible pinning force. In general, the mechanism underlying the anisotropy of the pinning forces remains to be investigated.

Lastly, the anisotropy of \vga\ in its SC state can be evidenced from a
peak splitting we observe in the $H^{\perp}$ configuration. As shown in
Fig.~\ref{fig:TF_muSR}, the FFT of the time-dependent {\textmu}SR asymmetry
shows a tiny, yet distinct peak splitting at low temperatures and low magnetic fields
for $\mu_0 H^{\perp} = 14$\,mT, which is absent for $\mu_0 H^{\parallel} = 14$\,mT.
Even when including a more robust apodization, a rectangular-like
peak still remains, indicative of a nontrivial spectrum, here
consisting of two near-lying frequencies [see Fig.~3~in Supplementary
Information]. Such splitting disappears and the two frequencies
merge either above a threshold temperature $\sim T_{c}/2$ (see
Fig.~\ref{fig:Bsplit}), or at a higher applied field (here,
$\mu_0 H^{\perp} = 50$\,mT, not shown). More importantly, in a parallel
magnetic-field configuration, no splitting is found (Fig.~\ref{fig:TF_muSR}a).
In view of a superconducting volume fraction of $\sim 100 \% $ and of
the very high quality of the investigated samples~\cite{Xiaofeng2024},
we can safely \emph{rule out any possible extrinsic origin} (like
inhomegeneities) of the peak splitting in the FFT of the time-dependent
asymmetry and consider it an intrinsic property of \vga. Whether such
key feature is due to anisotropy, to topological effects, or to
both remains open.\\

\noindent\textbf{Unconventional vortex lattice in \vga\ }\\[0.3ex]
The presence of a peak splitting in the local field distribution
$p(B)$, measured via \tfmu, is suggestive of the possible presence
of an unconventional vortex lattice, where two different length scales
coexist in a limited region of the $H$-$T$ phase diagram.
In this respect, Speight et al.\ have shown that nematic superconductors
with odd-parity pairing are expected to exhibit a non-Abrikosov-like
mixed state~\cite{Speight2024magnetic} consisting in stripes of
``coreless'' vortices, defined as two spatially separated half-quantum
vortices. Such unconventional structures were firstly named ``skyrmion''
vortices, because of their additional topological properties (namely,
a skyrmionic topological charge of $-1$)~\cite{Babaev2017}. This
unconventional mixed state is characterized by two length scales:
the separation between the chains
and the inter-skyrmion separation along the same chain. Depending on
the above two length scales, the local magnetic-field distribution $p(B)$
is expected to exhibit a characteristic double-peak structure. Such
feature, however, can only be observed in a small region of the
mixed state, at low temperatures and low applied fields. In fact,
for applied fields~\cite{Speight2024magnetic} or temperatures~\cite{TempPSplit}
greater than a threshold value, the skyrmion vortices are expected to
melt into a conventional Abrikosov vortex lattice. 

Although the skyrmionic vortex lattices can in principle be detected
by a series of techniques, 
such as scanning SQUID microscopy, scanning Hall probes, nuclear magnetic resonance, and small-angle neutron scattering~\cite{Speight2024magnetic}, only {\textmu}SR
has a sufficient signal-to-noise ratio, because muons are 100\% spin polarized.
Furthermore, unlike NMR and neutron scattering, which require magnetic
fields above $\sim 1$\,T, \musr\ can be performed at much lower fields,
for which the skyrmion vortex lattice --- if present --- is essentially unperturbed~\cite{Speight2024magnetic}.
More detailed \tfmu\ measurements at selected $H^{\perp}$-$T$ values
are required to characterize the nature of such an unusual mixed phase.
These studies are particularly relevant for \vga\ in view of the predicted
topological nature of its normal state~\cite{Xiaofeng2024}.

\vspace{10pt}
\noindent{\large\textbf{Conclusion}}\\
We performed local-probe (NMR and {\textmu}SR) investigations of the
normal- and superconducting-state properties of the binary compound \vga,
characterized by a topologically nontrivial normal state. Our measurements
reveal that, because of its pronounced 1D character due to vanadium chains,
\vga\ manifests a strong crystalline anisotropy on the electronic properties.
In the superconducting state, we find the superfluid density to be
\emph{fully-gapped,} with signatures of a double $s$-wave gap for applied
fields perpendicular to the \textit{c}-axis. The vanishing of the smaller
gap upon increasing the magnetic field intensity, or for applied fields
parallel to the $c$-axis, is consistent with a marked anisotropy of the
\vga\ superconductivity. The lack of spontaneous magnetic fields below
$T_c$ (see Supplementary Information) indicates a \emph{preserved} time-reversal symmetry in the
superconducting state. Finally, for $H^{\perp}=14$\,mT, the magnetic
field distribution in the mixed state presents a double-peaked FFT
structure, characteristic of the presence of an unconventional mixed phase.

In the normal state, NMR measurements reveal a strong anisotropy in the $H^{\parallel}$ and $H^{\perp}$ orientations, consistent with the 1D character of the compound. However, in
the measured temperature range, we did not find signatures of
topological nodes. Further measurements at higher temperatures
(i.e., at $k_\mathrm{B}T > \mu$, with $\mu$ denoting the chemical potential)
are needed to confirm the expected topological signatures in \vga.

\begin{figure}[!thp]
\centering
\includegraphics[width=0.49\textwidth]{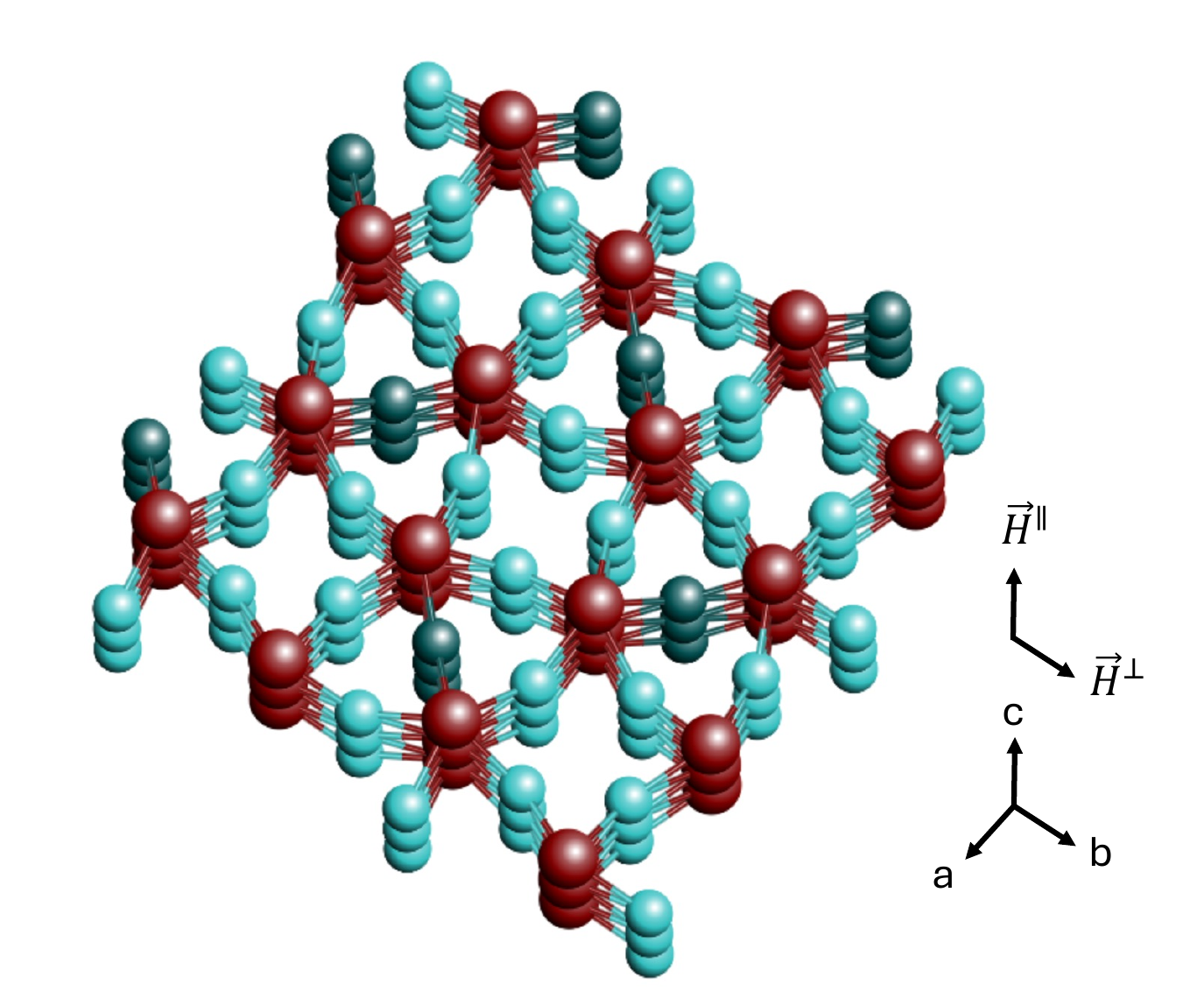}
\caption{\label{fig:structure}Crystal structure of \vga: light-green,
dark-green, and red spheres represent Ga(1), Ga(2) and V atoms, respectively.
The $c$-direction represents the main symmetry axis along which grow the 
needle-like single crystals, as shown in Figure~1d in Ref.~\onlinecite{Xiaofeng2024}. The black arrows indicate
the magnetic field directions. Here, $H^{\parallel}$ and $H^{\perp}$
represent the magnetic fields applied parallel- and perpendicular
to the $c$-axis, respectively.}
\end{figure}

\vspace{10pt}
\noindent{\large\textbf{Methods}}\\[0.3ex]
\noindent\textbf{Sample synthesis}\\[0.3ex]
Needle-like single crystals of \vga\ were grown from elemental
V in powder form in molten Ga flux. After inserting both elements into
an evacuated and sealed silica ampoule, a furnace was used to heat
the ampoule up to 1000$^\circ$C in 12\,h and to maintain it there
for another 12\,h.
The temperature was then dropped to 85$^{\circ}$C in 5\,h and kept there
for another 4 days. Finally, the ampoule was allowed to cool to
room temperature for 4 more days (see Ref. \onlinecite{Xiaofeng2024}
for further details). The resulting \vga\ crystals adopt a tetragonal
structure with a space group $P4/mbm$ (No.\ 127) and unit-cell parameters
$a = b = 8.9820(3)$\,\AA, and
$c = 2.7005(2)$\,\AA, in agreement with previous
work~\cite{Xiaofeng2024,Reddy1965,Lobring2002}. A schematic view
of the crystal structure is shown in Fig.~\ref{fig:structure}.\\

\noindent\textbf{DC magnetization}\\[0.3ex]
In recent studies, \vga\ was characterized by means of
resistivity (as a function of temperature and applied field), as well
as via heat-capacity, low-temperature thermal-conductivity
measurements~\cite{Xiaofeng2024} and \musr\ ~\cite{Cheng2024}.
Since our microscopic NMR and {\textmu}SR investigations were performed on samples of the same origin~\cite{Xiaofeng2024}, our basic sample characterization
included only complementary dc magnetometry measurements performed
on a MPMS2 SQUID magnetometer by Quantum Design. 
In our case, we covered an extended field range, up to
$\mu_{0}H_{\mathrm{app}} = 5$\,T, with the field applied
parallel and perpendicular to the $c$-axis ($H^{\parallel}$ and $H^{\perp}$, see Fig.~1~in the Supplementary Information).\\

\noindent\textbf{NMR experiments}\\[0.3ex]
To study the electronic properties of \vga\ with a complementary 
local probe, we
performed nuclear magnetic resonance (NMR) experiments. The sample was
oriented with the $c$-axis either parallel- or perpendicular to the external
magnetic field, $\mu_{0}H_{\mathrm{app}} = 5.007$\,T [See Fig.~1 in
supplementary information]. In principle, both gallium and
vanadium are suitable nuclei for such measurements. However, gallium
has two spin-3/2 nuclei, $^{69}$Ga and $^{71}$Ga, with  relatively large
quadrupole moments $Q$. 
Since, in a solid, the electric-field gradients at the nucleus
position can interact with the nuclear quadrupole moment, in case of a
large $Q$, this produces a considerable spectral broadening and complex overlapping peaks. 
On the other hand, despite a spin of 7/2, $^{51}$V is the only naturally
occurring isotope and has a large magnetogyric ratio while its
quadrupole moment is relatively small, $-0.043(5) \times 10^{-28}$\,m$^{2}$~\cite{Stone2016}.
Consequently, vanadium was the nucleus of choice for our NMR investigations.
%
%

Four main issues had to be overcome when performing $^{51}$V NMR.
Firstly, considering the suppression of $T_c$ in an applied
magnetic field (and the extensive {\textmu}SR study of the SC phase),
we had to limit the NMR measurements to the normal state of \vga.
Secondly, the NMR resonance frequency of $^{51}$V (here, 56.072\,MHz
from the reference compound VOCl$_3$), is close to that of
$^{63}$Cu (56.525\,MHz) and
$^{27}$Al (55.547\,MHz). Since $^{63}$Cu and $^{27}$Al are
ubiquitous among the NMR-probe materials, this leads to experimental
artefacts close to the $^{51}$V resonance line. During data processing, a
symmetrization procedure was performed to remove such residual 
artefacts that might still remain on both sides of the main $^{51}$V resonance. 
Thirdly, since a single \vga\ ``needle'' is too small to perform
NMR measurements, numerous needles ($\sim 30$) had to be bundled
together to obtain a sufficient mass for a suitable S/N ratio.
Fourthly, while it is straightforward to measure NMR in the $H^{\perp}$
configuration, where the \vga\ ``needles'' are parallel to the long axis of the coil,
the $H^{\parallel}$ configuration is significantly harder. To this aim a special sample bundle was realized (see Supplementary Information). Due to
practical problems with aligning the ``needles'' in the correct orientation,
as well as a poorer filling factor, the S/N ratio was significantly worse in this case. A large apodization
factor had to be used during post-processing, which distorts the
lineshape away from the main NMR carrier frequency. This fact, together
with the artefacts mentioned above, means that we could not accurately
determine the complete lineshape of the NMR resonance in the $H^{\parallel}$ configuration. Nevertheless, the line shifts
and the relaxation rates, where the apodization has a minimal effect,
could still be reliably obtained from the central transition, here
corresponding to the main NMR peak. \\

\noindent\textbf{Muon-spin rotation and relaxation ({\textmu}SR)}\\[0.3ex]
The bulk {\textmu}SR measurements were carried out at the multipurpose surface-muon spectrometer (Dolly) on the $\pi$E1 beamline and at the high-pressure muon facility (GPD) on the {\textmu}E1 beamline, both at the Swiss muon source of Paul Scherrer Institut, Villigen, Switzerland.
The low $T_c$ value required the use of a He-3 refrigerator,
essential for carrying out measurements in the 0.3--10\,K
temperature range in both facilities. 
For the surface-muon measurements, the samples were mounted on a
25-{\textmu}m thick copper foil, which ensured a good
thermalization at low temperatures. The needle-like samples were
oriented so that the $c$-axis (coincident with the needle axis)
was either parallel- or perpendicular to the main magnetic field.
Both transverse field (\tfmu) and zero field (\zfmu) muon-spin spectroscopy
measurements were performed in this case.

For the studies under pressure, a suitable sample consisting of
several crystals [see Fig.~2 in Supplementary Information] was inserted into
a double-walled MP35N/CuBe cylinder pressure cell (PC)~\cite{Khasanov2022,ShermadiniHPR,KhasanovHPR}, with the
$c$-axis parallel to the cell axis. Daphne oil 7373 was used as a pressure-transmitting medium to achieve hydrostatic conditions.
Here, low-temperature \tfmu\ measurements were performed at
about 0 and 19.5\,kbar, corresponding to a pressure of 2 and 22.5\,kbar
at room temperature, respectively.
The time-differential \musr\ data under normal- and high-pressure conditions were analyzed by means of the \texttt{musrfit} software package~\cite{Suter2012}. \\

\noindent{\large\textbf{Data availability}}

\noindent All the data needed to evaluate the reported conclusions 
are presented in the paper and/or in the Supplementary Information. 
Additional data related to this paper may be requested from the 
authors. The {\textmu}SR data were generated at the S{\textmu}S
(Paul Scherrer Institut, Switzerland).
Derived data supporting the results of this study are available from
the corresponding authors or beamline scientists. The \texttt{musrfit}
software package is available online free of charge at
http://lmu.web.psi.ch/musrfit/.
Computational results are available at Materials Cloud at  https://archive.materialscloud.org/record/2025.43


\vspace{10pt}
\noindent{\large\textbf{Acknowledgements}}\\
This work was supported by the
Schweizerische Nationalfonds zur F\"{o}r\-der\-ung der Wis\-sen\-schaft\-lichen
For\-schung (SNF) (Grant No.\ 200021\_169455). This work was partially
supported by the National Natural Science Foundation of China (Grants No.\
12274369 and 12304071). We ac\-knowl\-edge the allocation of beam time at the Swiss muon source (Dolly and GPD \musr\ spectrometers).
PB acknowledges computing resources provided by the STFC scientific
computing department's SCARF cluster and the CINECA projects CNHPC-1570115,
IsCa7\_CDWMKS and IsCb6\_TRSBKS. IJO acknowledges financial support from the PNRR MUR project ECS-00000033-ECOSISTER. This research was partially granted by University of Parma through the action Bando di Ateneo 2023 per la ricerca.\\

\vspace{10pt}
\noindent{\large\textbf{Author contributions}}\\
\musr\ experiments and data analysis: T.S., G.L., R.K. NMR experiments and data analysis: D.T., T.S. DC-magnetometry experiments and data analysis: G.L. Multidimensional order parameter theory: P.G. Sample growth and structural characterization: C.Q.X, X.Ke, X.Xu. DFT calculations: I.J.O, P.B. The manuscript was written by T.S., G.L., D.T., with input from all authors. Project planning and coordination: T.S.

\vspace{10pt}
\noindent{\large\textbf{Competing interests}}\\
The authors declare no competing financial or non-financial interests.

\vspace{10pt}
\noindent{\large\textbf{Correspondence}}\\
Correspondence and requests for materials should be addressed to X.X. or T.S.

\vspace{10pt}
\noindent{\large\textbf{Supplementary Information}}\\

\begin{figure}[t]
	\centering
	\includegraphics[width=0.23\textwidth,angle= 0]{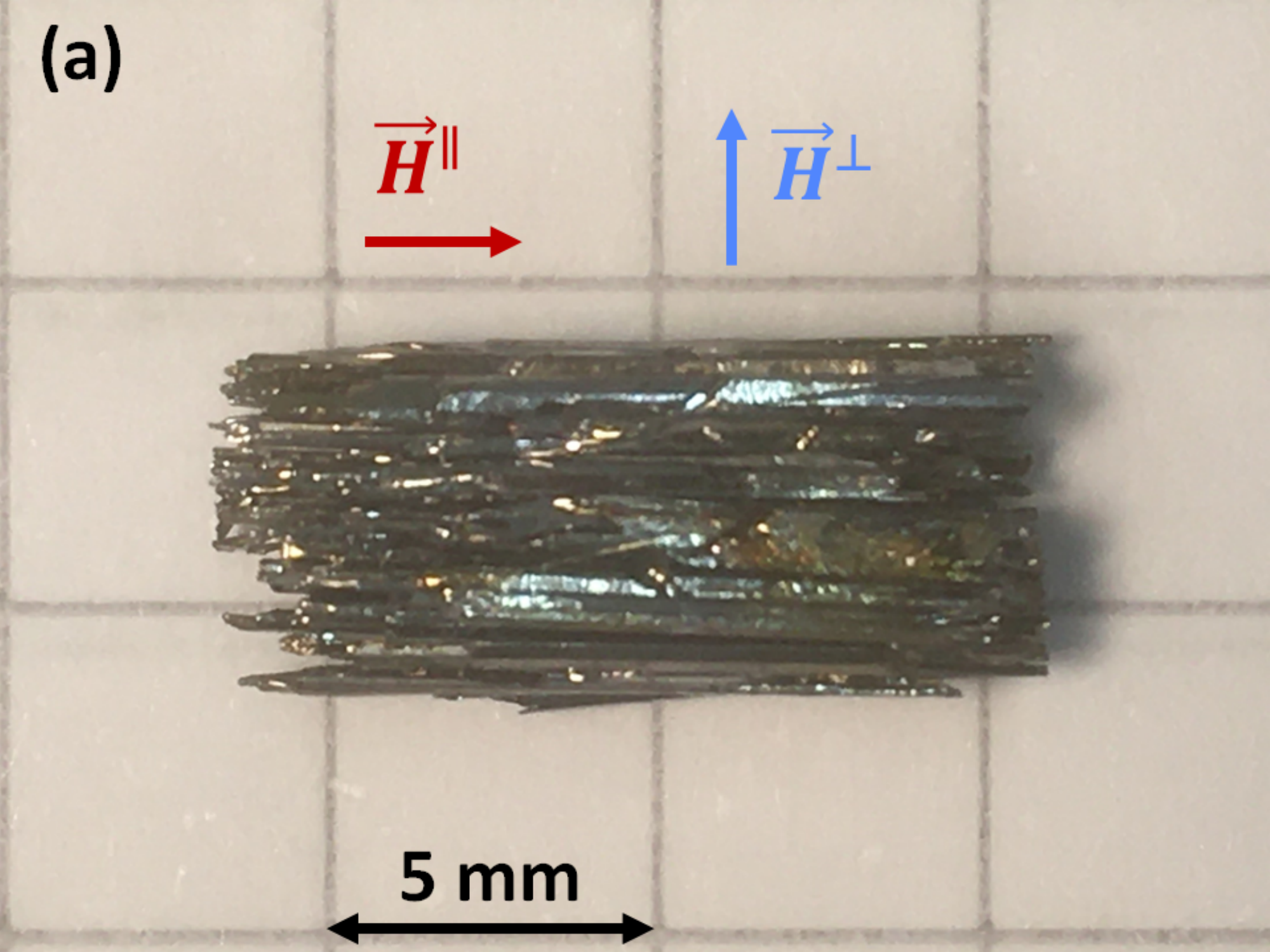}\hspace{2mm}
	\includegraphics[width=0.23\textwidth,angle= 0]{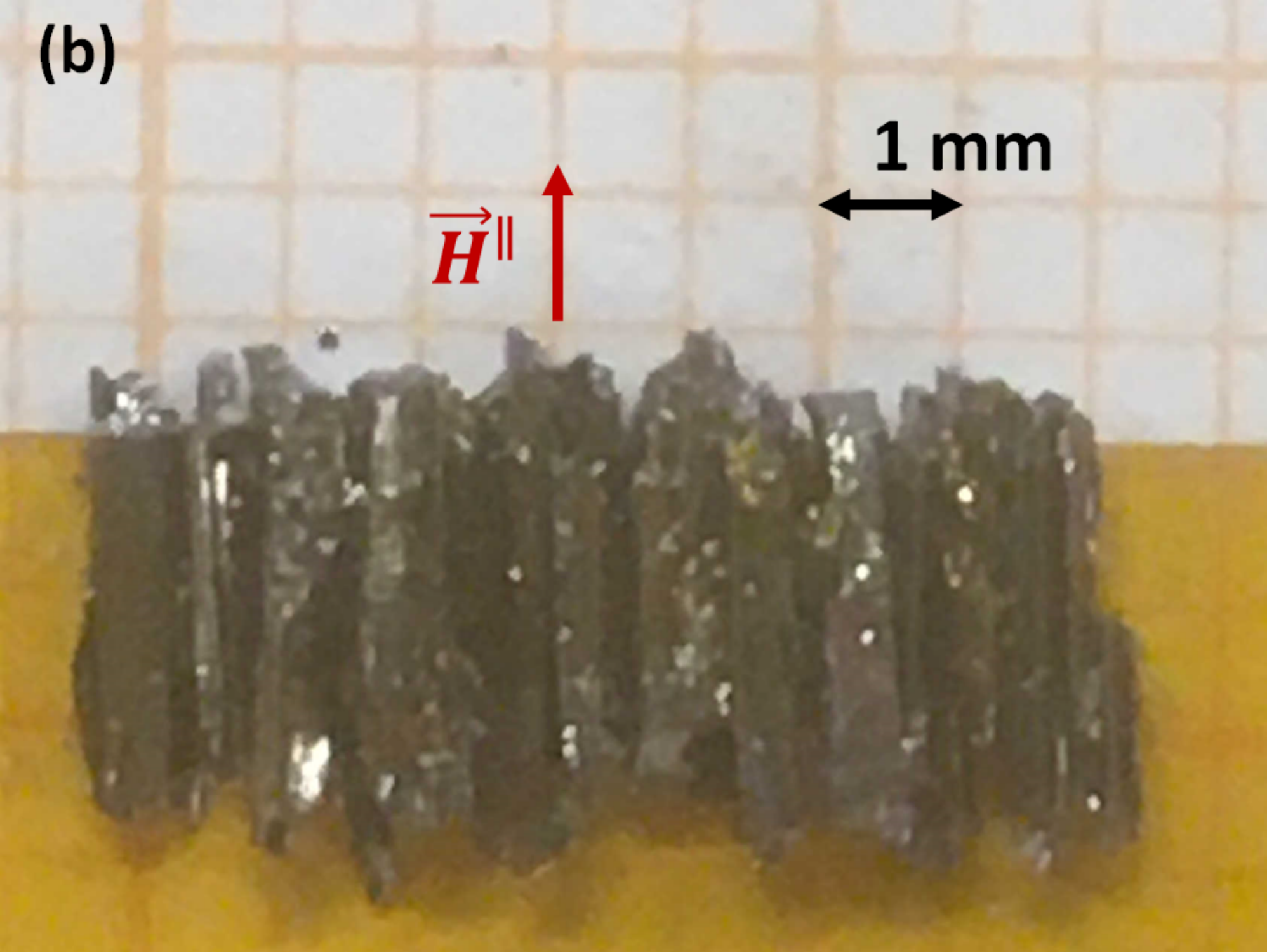}
	\caption{\label{fig:GPDsample}Bundles of \vga\ samples and applied
	magnetic field directions. The bundles are approximately cylindrical
	and of two types: (a) bundle with the longitudinal axis
	parallel to $c$-axis, (b) bundle with the longitudinal axis
	perpendicular to the $c$-axis.}
\end{figure}

\noindent\textbf{\textcolor{black}{Experimental setup}}\\[0.8ex]
The characteristic needle-like shaped \vga\ crystals require special
attention when selecting the most suitable configuration for applying
the magnetic field along the different 
crystal axes. For the dc magnetization measurements, the same single
crystal was used for both configurations, $H^{\parallel}$ and $H^{\perp}$
to the $c$-axis. In case of the \musr\ and NMR measurements, a single
crystal does not provide a sufficient signal-to-noise rato (SNR). Hence,
several crystals were glued together using GE varnish to form a cylindrical
bundle, 5.5\,mm in diameter and 12\,mm high, as shown in Fig.~\ref{fig:GPDsample}a. 
For the NMR measurements with $H^{\parallel}$ to the $c$-axis, another
bundle was prepared. This consisted of $\sim 2.5$\,mm long crystals glued
over a length of about $\sim 7.5$\,mm, as shown in Fig.~\ref{fig:GPDsample}b.
Here, the longest dimension coincides
with the NMR coil axis and is perpendicular to $c$.\\[1.5ex]

\noindent\textit{\textcolor{black}{$H^{\perp}$ to $c$-axis}}\\[0.5ex]
The \musr\ measurements under applied pressure (at the GPD instrument)
were performed in the $\mathrm{H}^{\perp}$ configuration using the sample
geometry given in Fig.~\ref{fig:GPDsample}a.
Here, the direction of the applied magnetic field, indicated by the blue arrow,
is perpendicular to the longitudinal axis of the sample bundle.
The NMR- and the standard {\textmu}SR measurements (at the Dolly instrument)
in the $\mathrm{H}^{\perp}$ configuration were also performed using
similar bundles.\\[1.5ex]

\noindent\textit{\textcolor{black}{$H^{\parallel}$ to $c$-axis}}\\[0.5ex]
For the standard \musr\ measurements in the $H^{\parallel}$ configuration
(again at the  Dolly instrument), the sample bundle of Fig.~\ref{fig:GPDsample}a
was rotated in plane by $90^{\circ}$.
In this geometry, the applied magnetic field (WEV magnet), indicated by
the red arrow, is parallel to both the $c$-axis and the longitudinal axis
of the bundle. In the case of NMR measurements in the $H^{\parallel}$
configuration, the sample dimensions are constrained by the diameter of
the RF coil. Here, the sample geometry given in Fig.~\ref{fig:GPDsample}b
had to be used, where the applied magnetic field, indicated by the red
arrow, is parallel to the $c$-axis but perpendicular to the longitudinal
axis of the sample bundle.\\[1.5ex]

%
\begin{figure}[!ht]
	\centering
	\vspace{-1mm}
	\includegraphics[width=0.40\textwidth,angle=0]{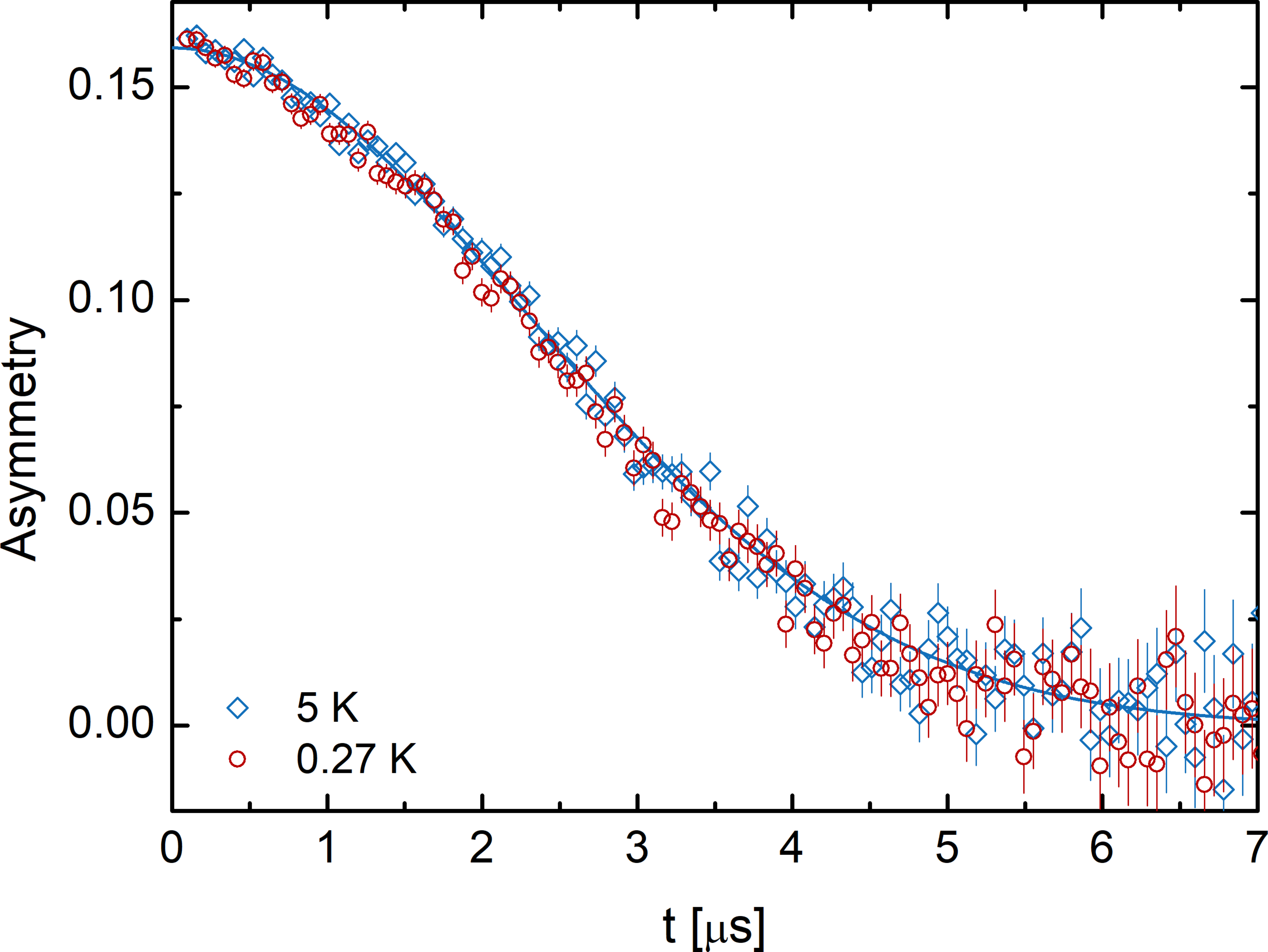}
	\caption{\label{fig:ZF-muSR}ZF-{\textmu}SR spectra collected in
	the superconducting- (0.27\,K) and the normal state (5\,K) 
	of \vga. The essentially overlapping datasets indicate the absence of any additional decay below $T_{c}$ (see text for details).}
\end{figure}

%
%
\noindent\textbf{\textcolor{black}{\zfmu\ measurements}}\\[0.8ex]
Among several superconducting states, odd parity superconductivity includes also  chiral, such as $p_{x}\pm ip_{y}$ \cite{Kallin2016}, or nonunitary spin-triplet states, both not preserving the time reversal symmetry (TRS). Since \vga\ is expected to be an odd parity superconductor it is important to verify if its superconducting state could preserve  the TRS. To this aim, we performed zero-field \zfmu\ measurements in the normal- and superconducting states. This technique is very sensitive to the weak spontaneous fields due to Cooper pairs in a spin-triplet state whose experimental evidence is an additional relaxation rate below $T_{c}$. Therefore \zfmu\ gives the possibility to unveil the breaking of TRS, as observed, e.g., in Sr$_2$RuO$_4$~\cite{Luke1998,Shiroka2012,Grinenko2021}, in Re-based superconductors~\cite{Singh2014,Shang2018a,Shang2018b,Shang2020ReMo,Shang2021a}, or in the putative topological superconductor Sr$_{0.1}$Bi$_2$Se$_3$~\cite{Biswas2019}.  Figure~\ref{fig:ZF-muSR} shows the time-dependent \zfmu\ muon spin polarization collected above- (5\,K) and below $T_c$ (0.27\,K). Such spectra can be modelled by a simple Gaussian-decay function: 
%
$P_\mathrm{ZF}(t) = \exp(-{\sigma_\mathrm{ZF}^{2}t^{2})/2}$, 
%
%
%
%
where $\sigma_\mathrm{ZF}$ represents the Gaussian relaxation rate. The resulting fit  parameters are $\sigma_\mathrm{ZF}(\mathrm{5~K}) = 0.441(5)$\,{\textmu}s$^{-1}$ ($\chi^{2}=1.035$) and $\sigma_\mathrm{ZF}(\mathrm{0.27\,K}) = 0.447(6)$\,{\textmu}s$^{-1}$ ($\chi^{2}=1.010$). The Gaussian relaxation rate model describes the static distribution of local fields at the muon implantation sites as generated by randomly oriented nuclear dipolar moments. Furthermore, both data set are fully superposed and no sizeable additional \musr\ relaxation is observed below $T_{c}$. Consequently, TRS is preserved in the SC state of \vga, which excludes the possibility of chiral superconductivity or of nonunitary spin-triplet states \cite{Kallin2016}. \\[1.5ex]
%

\begin{figure}[!h]
	\centering
	\vspace{-2mm}
	\includegraphics[width=0.41\textwidth,angle= 0]{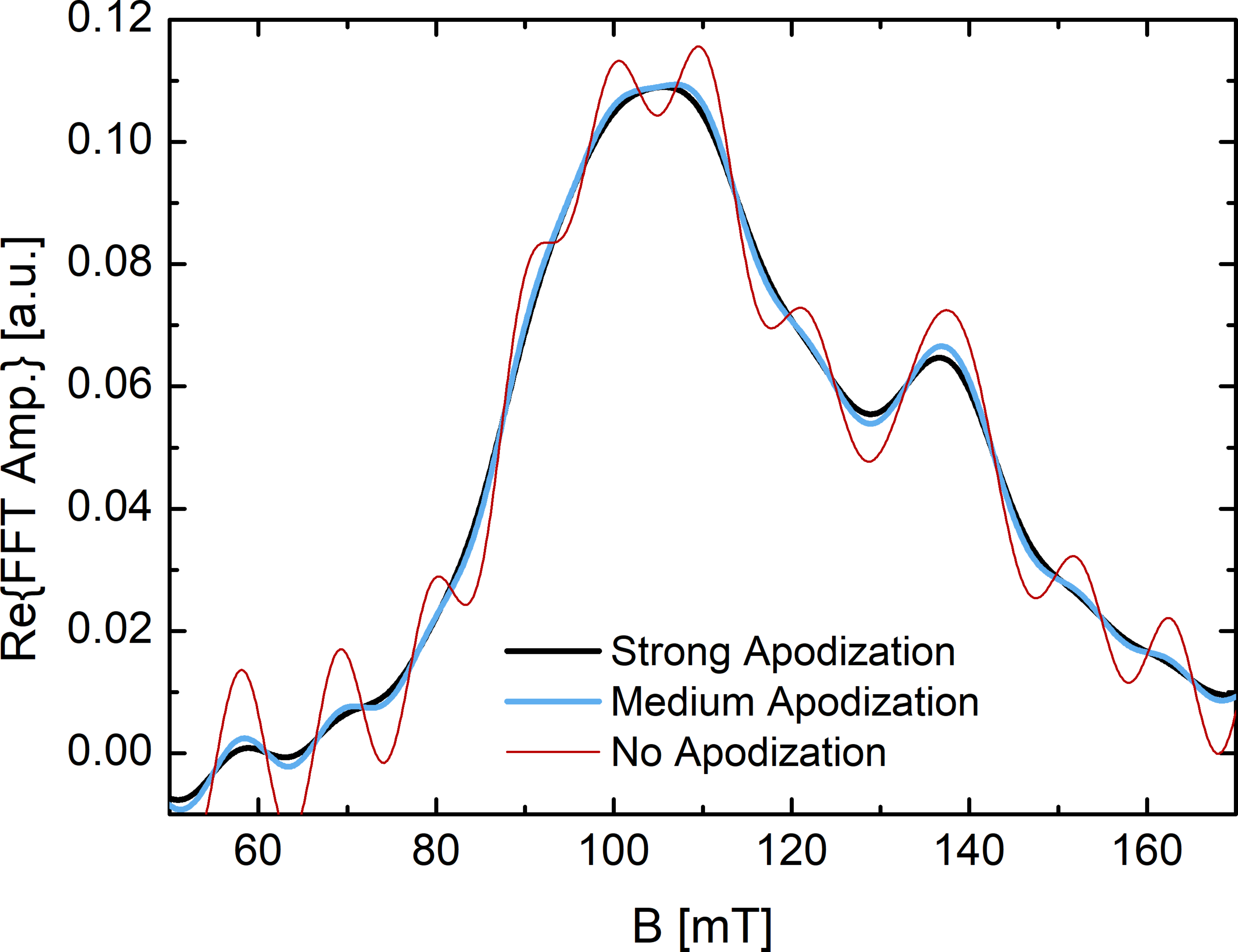}
	\caption{\label{fig:apodization}FFT of the time-dependent asymmetry,
	recorded at 0.9\,K at ambient pressure in the orthogonal
	configuration, here shown with: no-, medium-, and strong apodization.
	Note the persistence of the flat-top peak, independent of the used apodization.}
\end{figure}

\noindent\textbf{\textcolor{black}{Apodization}}\\[0.8ex]
To corroborate the splitting into two peaks of the superconducting component of FFT below $T_{c}/2$, in Fig.~\ref{fig:apodization} we plot the FFT of the
time-dependent asymmetry recorded at 0.9\,K with no, with medium,
and with strong apodization. The clear (but noisy) peak splitting
in the absence of apodization becomes a flat-top peak under apodization,
thus suggesting the contribution of two close-lying (though now
unresolved) frequencies.
Since the two-peak feature is robust (i.e., independent of the apodization), this excludes that it arises from a data-processing artifact and
confirms its real nature.\\[1.5ex]


\vspace{10pt}
\noindent{\large\textbf{References}}
\bibliographystyle{naturemag}
\bibliography{V2Ga5_v13}
\end{document}